\documentclass[nofootinbib,aps,pra,reprint,twocolumn,groupedaddress,nofootinbib,10pt]{revtex4}

\usepackage[T1]{fontenc}
\usepackage{bm}
\usepackage{amsfonts}
\usepackage{graphicx}
\usepackage{slashed}
\usepackage{color}
\usepackage{soul}
\usepackage{cancel}
\usepackage{hyperref}
\usepackage{amsthm}
\usepackage{amssymb}
\usepackage{amsmath}
\usepackage{xparse}
\usepackage{xspace}
\usepackage{mathtools}
\usepackage{etextools}
\usepackage{ifthen}
\usepackage{empheq}
\usepackage[most]{tcolorbox}
\newcommand{\Tends}{\,\rightarrow\, }
\newcommand{\COM}[2]{\ensuremath{\left[{#1} ,{#2}\right]}\xspace}
\newcommand{\Order}[1]{\mathcal{O}\left(#1 \right)}

\DeclarePairedDelimiter{\MbBraces}{\{\!\{}{\}\!\}}
\newcommand{\PB}[2]{\left\{#1,#2\right\}}
\newcommand{\QPB}[2]{\left\{#1,#2\right\}}
\newcommand{\MB}[2]{\MbBraces{#1,#2}}
\newcommand{\MMB}[2]{\PB{#1}{#2}_\mathcal{W}}
\newcommand{\ud}{\mathrm{d}}

\newcommand{\uHam}{\mathbf{\Ham}}
\newcommand{\uW}{\mathbf{W}}
\newcommand{\uF}{\mathbf{F}}
\newcommand{\uG}{\mathbf{G}}

\newcommand{\uK}{\mathbf{K}}
\newcommand{\uH}{\mathbf{H}}

\newcommand{\HPD}[3]{\ensuremath{\frac{\partial^{#3}#1}{{\partial#2}^{#3}}}\xspace}
\NewDocumentCommand{\PD}{o m m}{%
  \IfNoValueTF{#1}
    {\ensuremath{\frac{\partial #2}{\partial #3}}}%
    {\ensuremath{\frac{\partial^{#1} #2}{\partial #3^{#1}}}}%
  \xspace
}
\NewDocumentCommand{\TD}{o m m}{%
  \IfNoValueTF{#1}
    {\ensuremath{\frac{\ud #2}{\ud #3}}}%
    {\ensuremath{\frac{\ud^{#1} #2}{\ud #3^{#1}}}}%
  \xspace
}
\newcommand{\ui}{\mathrm{i}}

\newcommand{\Lgn}{\mathcal{L}}
\newcommand{\Ham}{{H}}

\begin{document}
\enlargethispage{300pt}

\title{From Mass-Shell Factorisation to Spin: An Attempt at a Matrix-Valued Liouville Framework for Relativistic Classical and Quantum Phase-Spacetime}

\author{Mark Everitt}
\affiliation{Quantalytics, Loughborough, UK}
\email{m.j.everitt@physics.org}
\date{\today}

\begin{abstract}
Here we argue that spinor structure arises naturally if relativistic statistical mechanics is formulated directly on phase spacetime. 
Requiring a first-order phase-spacetime description that retains both mass-shell branches leads to a Clifford factorisation of the relativistic constraint and hence to a $4\times4$ spinor-matrix distribution function.
We show that deformation quantisation leads to a phase-space formulation of spin quantum mechanics.
We argue that projection onto positive- and negative-energy sectors recovers the standard relativistic classical transport equations in the appropriate scalar limits, while the corresponding left- and right- stargenvalue equations reproduce the constraint structure of the Dirac-Wigner formulation.
The result is a phase-space route from relativistic statistical mechanics to spinor quantum mechanics, in which spin algebra emerges as the internal structure
required by any relativistic statistical theory containing both mass-shell branches and the dimensions of angular momentum from quantum non-locality.
\end{abstract}

\maketitle

\tableofcontents

\section{Motivation}
From a philosophical positivist perspective it is valid to argue that empirically science should always be framed as a probabilistic theory. 
There will always be intrinsic uncertainties in initial conditions, physical parameters, and measured values. 
Even if one does not hold this view, it seems natural to argue that any physical theory should allow statistical descriptions of ensembles.
We could therefore argue that the requirement of a probabilistic/statistical representation of a system is as fundemantal as the foundational principles of least (extremal) action, Noether's Theorem (linking symmetries to conservation laws) and relativity (that the predictions of a theory should not be dependent on the `observer').
If we determine within a theory that there is an incompatibility between any of these principles then that is a cause for concern and investigation.

Consequently, the Hamiltonian formulation of Liouville’s theorem merits particular consideration where, for example \emph{``we have to insist that, in a covariant context, there is no Hamiltonian with the meaning of an energy''}~\cite{hakim2011introduction}. 
This is compounded further by the Currie–Jordan–Sudarshan theorem, which shows that, for a finite number of classical point particles with canonical coordinates whose positions transform covariantly, there is no non-trivial Hamiltonian interaction compatible with the full Poincaré symmetry (the fundamental symmetries of special relativity)~\cite{RevModPhys.35.350}. 
That said, classical field theories such as electromagnetism can be perfectly relativistically covariant. 
One may view this as strong motivation for the field-theoretic viewpoint of relativistic quantum field theory, and it helps to explain why relativistic one-particle wave equations such as the Dirac equation appear conceptually less problematic than a fully covariant Hamiltonian formulation of the relativistic Liouville equation. 
Moreover, the fact that relativistic Liouville is a statistical theory that can be cast as a field theory on phase space provides a first indication that a suitably modified Liouville equation might resolve some of these difficulties.

This perspective suggests revisiting phase-space methods as a natural setting in which to compare classical and quantum descriptions on as equal a footing as possible.
From a pedagogical viewpoint, there is considerable value in teaching quantum physics through its phase-space formulation, wherein quantum mechanics naturally emerges as a deformation quantization of classical physics, replacing the Poisson bracket with the Moyal bracket and conventional distribution multiplication with the star product. 
However, a persistent gap in this approach has been the difficulty of naturally introducing spin, since quantum mechanical spin-$1 \over 2$ has no direct analogue in classical mechanics. 
Despite significant prior efforts using spin phase-space representations via $\mathrm{SU}(N)$ Wigner functions formulated in terms of a displaced-parity operator, a satisfying quantum-classical analogy entirely within the language of deformation quantisation has remained elusive. 

Motivated by these difficulties in incorporating spin-$1 \over 2$ into a phase-space and deformation-quantisation framework, I was led to revisit the standard derivation of the Dirac equation. 
Of particular interest were, on the one hand, the arguments leading to the Clifford-algebraic structure that makes relativistic spin natural and, on the other, the similarities between the quantum master (von Neumann) equation, the Moyal equation and the Hamiltonian formulation of Liouville’s theorem. 
The strategy developed below revisits Dirac’s original derivation, but reformulates it in a classical phase-space language: 
the relativistic Liouville equation replaces Schr\"odinger’s equation as the dynamical starting point, and spin degrees of freedom are encoded via a Clifford algebra–valued distribution. 
For clarity, the analysis is restricted to a single particle in free space or minimally coupled to an external electromagnetic field. 
For a relativistic extension of Schr\"odinger’s equation, Dirac argued that 
(a) it should make use of the mass shell, 
(b) the dynamical equation, like the Schr\"odinger equation, should be first-order in time but also, inspired by issues with the Klein–Gordon equation, that the Hamiltonian should be first-order in space, and 
(c) the four-momentum components should be replaced with their operator counterparts. 
Unlike Dirac’s argument, I will use the Liouville equation as the basis for the discussion, but apart from this, and the fact that no operator substitution is required, assumptions (a) and (b) will be retained.
We will then argue for a matrix relativistic Liouville equation and a matrix probability density that we will term a spinor-matrix distribution function before investigating deformation quantisation of the resulting model. 

\paragraph*{Organisation of the paper:}

This paper proceeds in four stages. 
First, Sections~II and~III develop the relativistic statistical argument. 
Section~II can be safely skipped by those familiar with the analytical mechanics of the relativistic free particle as its primary purpose is to make the document self-contained and to define notation.
After fixing the free-particle notation, In section~III, we show that retaining both mass-shell branches in a first-order phase-spacetime description naturally leads to a Clifford-valued, spinor-matrix distribution. 
Projection onto the energy sectors is then used as a consistency check, recovering the usual relativistic Liouville transport equations in the diagonal scalar limits.

Second, Section~IV applies the construction to a charged particle in an external electromagnetic field. 
This acts as a test of both the ordered and Weyl-symmetrised matrix transport equations and shows how the standard spinless Liouville-Vlasov equation is recovered when the off-diagonal sector structure is absent.

Third and finally, Section~V applies deformation quantisation to the approach. 
We showw that the matrix star product introduces the non-commutative and non-local phase-space structure needed for spin quantum mechanics, while the corresponding left and right stargenvalue equations connect the framework to the Dirac-Wigner formulation.

\section{Background}
As a starting point for our argument, recall that in classical mechanics a single 
particle with Hamiltonian $H(\mathbf{x},\mathbf{p},t)$ has a phase-space probability distribution $\rho(\mathbf{x},\mathbf{p},t)$ that evolves according to the Liouville equation
\begin{equation}
  \frac{\partial \rho}{\partial t} + \{\rho,H\} _3 = 0,
\label{eq:liouville}
\end{equation}
where $\{A,B\}_3$ is the usual Poisson bracket in three dimensional space. 
Importantly note that, as with the Schr\"odinger equation,
eq.~\eqref{eq:liouville} is \emph{first-order} in time and, as for the Dirac constraints, the Poisson bracket is first order in phase-space variables. Note that a core assumption of \textbf{any statistical theory} is that the representation of a state at any given time  be sufficient to dictate all future evolution and thus any governing equation for the evolution of the state must be a differential equation that is at most first order in time.

The Dirac equation resulted from seeking a relativistic generalization of Schr\"odinger's equation.
We will therefore seek relativistic generalizations of this equation with the following conditions:
\begin{enumerate}
	\item[C1] First-order in space and time [instead we might equivalently require that our dynamical framework does  (i) not  reduce the dimension of the phase-space or (ii) exclude any solutions - i.e. contains both positive and negative energy solution].
	\item[C2] Consistent with the mass-shell condition.
\end{enumerate} 
The key difference from the argument for the Dirac equation is that (i) we will not replace energy and time by their operator counterparts; (ii) by first-order in space we mean all phase-space derivatives.
The relativistic Liouville equation is given by the extended relativistic Poisson bracket vanishing~\cite{MARSDEN198629,hakim2011introduction}
\begin{equation}
	\PD{\Ham(x,p)}{p_\nu}\PD{W(x,p)}{x^\nu} -
	\PD{W(x,p)}{p_\nu}\PD{\Ham(x,p)}{x^\nu} \equiv \PB{\Ham}{W} = 0. \label{eq:relL}
\end{equation}
As with the non-relativistic Liouville equation this is also consistent with our constraint of being first order in phase-spacetime variables. 
At this stage one might consider the scalar relativistic Liouville equation sufficient. The motivation for the present work is that treating both mass-shell branches within a single first-order phase-spacetime object suggests a linear factorisation of the constraint. 
This leads naturally to a Clifford-valued distribution rather than to a scalar branch-labelled density.

\subsection{Relativistic free particle: momentum \& Hamiltonian}
We will provide a derivation of the momentum and free Hamiltonian following similar lines to standard textbook arguments~\cite{Landau:1975pou}.
We will however do so in a way that makes clear and minimises assumptions. 
This adds some length to the discussion but does serve to make sure all notation is well defined and this manuscript is reasonably self-contained. 
As usual, coordinates are defined as $$x^\mu=(x^0,x^1,x^2,x^3)=(ct,x^1,x^2,x^3).$$ We follow the convention that Greek indices run over all spacetime components and Latin indices are restricted to spatial components.
Proper time $\tau$ is defined along any given worldline by
\begin{equation}
    \ud s = c\,\ud\tau.
\end{equation}
where the line element $\ud s$ is defined via the Minkowski metric [with the usual signature $(+,-,-,-)$] so that
\begin{equation}
    \ud s^2 = \eta_{\mu\nu} \ud x^\mu \ud x^\nu
            = c^2 \ud t^2 - \ud\Vec{x}\,^2.
\end{equation}
The three velocity is $v^i \equiv \ud x^i/\ud t $ which we will only ever write with raised index (or as the geometric vector $\Vec{v}$) because the three velocity $\Vec{v}$ is fundamentally contravariant in its components. 
 Noting that 
\begin{equation}
    c^2 \ud\tau^2 
    = 
    c^2 \ud t^2\left(1 - \frac{v^2}{c^2}\right)
\end{equation}
we see that coordinate time is connected to proper time by
\begin{equation}
    \TD{t}{\tau} 
    = 
    \gamma \equiv \frac{1}{\sqrt{1 - v^2/c^2}}.
\end{equation}
The four-velocity is
\begin{align}
    u^\mu &\equiv \TD{x^\mu}{\tau}
    = 
    \left(c\,\TD{t}{\tau}, \TD{x^i}{t}\TD{t}{\tau}\right)
    \\ &= 
    \gamma\left( c,  \Vec{v}\right)
    \\
    u_\mu  & =\gamma\left( c,-  \Vec{v}\right).
\end{align}
The invariant action for a free point particle of mass $m$ is:
\begin{equation}
    S[x] = -m c \int \ud s,
\end{equation}
where the `$-$' sign of the action is chosen to ensure that the usual non-relativistic limit is recovered\footnote{The canonical approach can be generalised to  work in any coordinate system and even for curved spacetime by replacement of the Minkowski metric and modifying specific formulae from that metric. 
See~\cite{Horwitz_2019} for example which also discusses the Stueckelberg, Horwitz and Piron~(SHP) approach to canonical classical and quantum
dynamics for general relativity and shows ``canonical Poisson brackets of the SHP theory
remain valid (invariant under local coordinate transformations) on the manifold of GR, and
provide the basis, following Dirac’s quantization procedure [replacing Poisson brackets with commutation relations], for formulating a quantum theory.
The theory is developed both for one and many particles.''. 
We will later take a different approach and look at direct deformation quantisation of the phase space itself. 
Another reason for noting the generalisability of this approach beyond special relativity is that it may later provide a path to generalise the work presented here to curved space-times.}.

Returning to the main argument, 
We will \textbf{assume} a time-like world-line so that $\ud s^2>0$ and therefore:
\begin{equation}
    \ud s = \sqrt{\eta_{\mu\nu}\,\ud x^\mu \ud x^\nu}.
\end{equation}
If $\lambda$ is an arbitrary parameter along a given worldline $x^\mu(\lambda)$ we define\footnote{Here and below, a dot always denotes derivative with respect to the arbitrary parameter $\lambda$. Derivatives with respect to coordinate time $t$ and proper time $\tau$ will be written explicitly.}
\begin{equation}
    \dot x^\mu \equiv \frac{dx^\mu}{d\lambda}.
\end{equation}
Then
\begin{equation}
    \ud s 
    = 
    c \ud \tau 
    =
    \sqrt{\eta_{\mu\nu}\,\dot x^\mu \dot x^\nu}\, \ud\lambda.
\end{equation}
We will note for later use that this yields
\begin{equation}\label{eq:dtaudl}
    \TD{\lambda}{\tau} 
    = 
    \frac{c}{ \sqrt{\eta_{\mu\nu}\,\dot x^\mu \dot x^\nu}}.
\end{equation}
The action becomes
\begin{equation}
    S[x] 
    = 
    -m c \int d\lambda\;\sqrt{\eta_{\mu\nu}\,\dot x^\mu \dot x^\nu},
\end{equation}
and the Lagrangian is
\begin{equation}
    \Lgn(x,\dot x) 
    = 
    -m c\,\sqrt{\eta_{\mu\nu}\,\dot x^\mu \dot x^\nu}.
\end{equation}
By definition, the canonical momentum conjugate to the coordinate $x^\mu$ is
\begin{equation}
    \pi_\mu \equiv \PD{\Lgn}{\dot x^\mu}.
\end{equation}
so that 
\begin{align}
    \pi_\mu
    &= 
    \PD{}{ \dot x^\mu}
    \left[-m c 
    \sqrt{
        \eta_{\alpha\beta}\,\dot x^\alpha \dot x^\beta
        }\right] \nonumber\\
    &= 
    -m c\, \frac{1}{2}
    \left(\eta_{\alpha\beta}\,\dot x^\alpha \dot x^\beta\right)^{-1/2}
    \cdot
    2\,\eta_{\mu\nu}\,\dot x^\nu \nonumber\\
    &= 
    -m c\,
   \frac{\eta_{\mu\nu}\,\dot x^\nu}
        {\sqrt{\eta_{\alpha\beta}\,\dot x^\alpha \dot x^\beta}}.
\label{eq:p_from_L}
\end{align}
Note that the factor of two arises in the sum over repeated indices, i.e. because $\eta_{\alpha\beta}\,\dot x^\alpha \dot x^\beta$ has a quadratic nature. 
Using the chain rule on the four velocity and using equation~\ref{eq:dtaudl}
\begin{equation}
    u^\mu
    = 
    \frac{dx^\mu}{d\lambda}\,\frac{d\lambda}{d\tau}
    = 
    \dot x^\mu \,\frac{d\lambda}{d\tau}
    = 
    \dot x^\mu \,\frac{c}{\sqrt{\eta_{\alpha\beta}\,\dot x^\alpha \dot x^\beta}}.
\end{equation}
premultiplying by $\eta_{\mu\nu}$ and rearranging,
\begin{equation}
    \frac{\eta_{\mu\nu}\,\dot x^\nu}{\sqrt{\eta_{\alpha\beta}\,\dot x^\alpha \dot x^\beta}}
    =
    \frac{1}{c}\,\eta_{\mu\nu} u^\nu =\frac{1}{c} u_\mu.
\end{equation}
Substituting this into Eq.~\eqref{eq:p_from_L} yields:
\begin{equation}
    \pi_\mu =  -m u_\mu = -m \gamma (c,-\Vec{v})
\end{equation}
Specifically this $\pi_\mu$ is is the canonical momentum in the cotangent bundle and is a result of the canonical definition $\pi_\mu = \partial L / \partial \dot x^\mu$ applied to the invariant free-particle Lagrangian.

Now because 
$$
    \pi^\mu  =-\gamma m ( c, \Vec{v})
$$
the physical four momentum is often defined as
\begin{equation}
    p^\mu = -\pi^\mu =\gamma m( c, \Vec{v})
\label{eq:phyical_momentum}    
\end{equation}
so that the zeroth component (energy) is positive (which, as shown in~\cite{Landau:1975pou} is also the momentum  $p_\mu = \frac{\partial S}{\partial q^\mu}$ taking the action as Hamilton's principal function -- both are the same Noether charge up to an arbitrary overall sign). 

\subsection{Relation to the usual Hamiltonian $H$}

If we choose the parameter $\lambda = t$ (coordinate time), the Lagrangian becomes
\begin{equation}
    L(\mathbf{x},\mathbf{v}) 
    = 
    -m c^2\sqrt{1 - \frac{v^2}{c^2}}.
\end{equation}
The canonical three-momentum $\Vec{p}$ is then defined by
\begin{equation}
    p^i 
    \equiv  
    \frac{\partial L}{\partial v^i}
    = 
    \gamma m v^i,
\end{equation}
which matches the spatial components of $p^\mu$.

The Hamiltonian is
\begin{equation}
    H 
    = 
    \Vec{p}\cdot\Vec{v} - L
    = 
    \gamma m v^2 + m c^2\sqrt{1 - \frac{v^2}{c^2}}.
\end{equation}
With a little algebra, this simplifies to
\begin{equation}
    H = \gamma m c^2.
\end{equation}
Comparing with Eq.~\ref{eq:phyical_momentum} we see
\begin{equation}
    p_0 = p^0 =  \frac{H}{c},
\end{equation}
again as a derived identity that follows from
(a) the canonical definition of momentum,
  (b) the specific invariant free-particle Lagrangian.

The central claim of this work is that spinor structure need not be introduced as an independent quantum postulate. 
If a relativistic statistical phase-space theory is required to remain first order, to retain both mass-shell branches, and to encode the factorised mass-shell constraint linearly, then the natural solution  is to use a Clifford algebra. 
In space-time this is satisfied by the four-component Dirac spinor space. 
The projected positive- and negative-energy sectors are rank-two subspaces, and the remaining two-dimensional internal structure is precisely the spinor degree of freedom. 
In this sense, the algebra of spin arises kinematically from the compatibility of relativistic mass-shell factorisation with a statistical phase-space description.
The route to relativistic quantum mechanics considered here is then Moyal deformation quantisation. In this step, $\hbar$ supplies, through non-locality, the physical scale of angular momentum associated with the internal spin algebra.
In the classical limit $\hbar \Tends 0$, the Clifford spin algebra will remain as an internal algebraic structure, while the angular-momentum scale associated with `spin' vanishes with $\hbar$.
\section{Free particle Relativistic Liouville equation and the spinor-matrix distribution function}
\subsection{Candidate Hamiltonians and Governing Equations}
Because the relativistic Liouville equation is formulated on phase space with canonical variables we cannot express the  Hamiltoian as $H=\gamma mc^2$ we must
instead express it in terms of the canonical position and momentum.
From the mass shell condition 
$$
    p^\mu p_\mu = p_0^2 - p_x^2-p_y^2-p_z^2 
    =
    m^2c^2.
$$
it is usual to write the Hamiltonian as $$E_p=\sqrt{ c^2 \Vec{p}\,^2+m^2c^4}$$ but this excludes negative energy solutions and violates condition C1.
To find a Hamiltonian consistent with C1 and C2 we will follow Dirac and seek some constants $\{\gamma^\mu\}$  such that
$$
    \gamma^\mu p_\mu 
    = 
    \gamma^0p_0 + \gamma^1 p_x+ \gamma^2 p_y+ \gamma^3 p_z
    = 
    m c.
$$
In $3+1$ dimensions, the minimal Clifford algebra representation has 
dimension $4\times4$. 
If one wants to factorize the above expression a linear \emph{4-component} structure is forced. 
The solution that Dirac found in relativistic quantum mechanics, where these $\gamma^\mu$ are the same $\gamma$-matrices of that work is also a solution here (the $p_i$ do not need to be operators to force the same matrix outcome).
Inspired by the Dirac equation we might look for an equation of the form 
\begin{equation}
    \left(\gamma^\mu p_\mu - mc 1_4 \right) \psi = 0_4. \label{eq:df}
\end{equation}
Writing $p_{0}=\Ham/c$ and multiplying on the left by $\gamma^{0}$ gives
\begin{align}
    \left(\frac{\Ham}{c} 
    - \Vec{\alpha}\cdot \Vec{p} 
    - \gamma^0 m c \right) \psi
    =0_4,	
\end{align}
with
\begin{align}
    \alpha^{i} 
    &= 
    \gamma^{0}\gamma^{i}, 
    \\
    \COM{\alpha^{i}}{\alpha^{j}}_+
    &=2
    \delta^{ij},
    \\
    \COM{\alpha^{i}}{\gamma^0}_+&=0_4.
\end{align}
where $\COM{\cdot}{\cdot}_+$ is the anti-commutator. Let us note for later reference that $\gamma^0$ is sometime given the following notation
$$
    \gamma^0
    =
    \beta\mathrm{,\ so\ } \gamma^i=\beta \alpha^i.
$$
The matrix-valued Hamiltonian
\begin{equation}
\boxed{
    \uHam_\mathrm{3+1} = c\,\alpha^ip_i + \gamma^0 m c^2
}
\end{equation}
This Matrix Hamiltonian is however defined in terms of the three spatial momenta only and is not covariant (and so we have chosen the label $3+1$ to reflect this). Our argument has gone too far. 

Let us retain the idea of multiplying by $c \gamma^0$ but return to the Dirac factorisation and define a matrix super-Hamiltonian
\begin{equation}
    \boxed{
    \uH_\mathrm{free}(x,p)= c \gamma^0 \left(\gamma^\mu p_\mu - mc 1_4 \right)
    }
\end{equation}
which is  first order in every momentum component. While multiplication by $\gamma^0$ prevents this form from being covariant its adoption results in a physically motivated argument that reproduces the usual positive and negative energy Liouville dynamics by projection onto the appropriate energy surfaces. A fully covariant approach using
\begin{equation}
    \uH_\mathrm{free}^\mathrm{cov}(x,p)= c \left(\gamma^\mu p_\mu - mc 1_4 \right) \label{eq:HFreeCov}
\end{equation}
is also possible and mathematically more elegant. A summary of that approach is presented in~\ref{sec:cov}.

We now have a Hamiltonian that we can use to test the approach by  
substitution into a relativistic Liouville equation of the form
\begin{align}
	\PD{\uH(x,p)}{p_\nu}\PD{\uW(x,p)}{x^\nu} -
	\PD{\uW(x,p)}{p_\nu} \PD{\uH(x,p)}{x^\nu}   
	 \equiv \PB{\uH}{\uW} = 0_4, \label{MLE}
\end{align}
where, to make sense $\uW$ must be a $4 \times 4$ matrix because it must live in the same non-commutative algebra as the Hamiltonian, $\uH(x,p)$, that drives its dynamics.
In sumarary: to keep the relativistic Liouville equation linear in derivatives, and treat both energy signs at once, $\uHam$ must be matrix-valued. 
Once $\uHam$ lives in the Clifford algebra, consistency requires $\uW$ to live in the \emph{same} algebra so that the Poisson bracket $\{\uHam,\uW\}$ is well defined.
Moreover, Lorentz rotation or boosts acts/transforms in the expected way $\mathbf{S}(\Lambda)\uW \mathbf{S}^{-1}(\Lambda)$.
We will refer to $\uW$ as a spinor-matrix distribution function.

Because $\uH$ and $\uW$ are now matrix-valued, the order of
factors in the bracket matters.  The ordered bracket used here is therefore a proposed first-order matrix-valued evolution equation, rather than as a genuine Poisson bracket on a
matrix algebra. While this bracket satisfies two of the axioms of a Lie algebra (bilinear and  alternating) and it fails the Jacobi identity and Leibniz rule and is therefore not a derivation with respect to matrix multiplication. Exploring the importance of this observation is beyond the scope of this initial exploratory work.

Now, as $\uH(x,p)$ and $\uW(x,p)$ no longer commute there are alternative brackets that would reproduce the normal Poisson bracket if $\uH$ and $\uW$ commute. For example consider the symmetric (Weyl ordered) bracket 
\begin{align}
    &
    \MMB{\uH}{\uW}	\equiv \label{extendedbracket} \\ 
    &
    \nonumber \left(
	\PD{\uH}{p_\nu}
	\PD{\uW}{x^\nu}
	+
	\PD{\uW}{x^\nu}
	\PD{\uH}{p_\nu}
	\right)
    -
	\left(
	\PD{\uH}{x^\nu}
	\PD{\uW}{p_\nu}
	+
	\PD{\uW}{p_\nu}
	\PD{\uH}{x^\nu}
	\right)
\end{align}
which (up to an irrelevant factor of $2$) leads to an alternative to the  relativistic Liouville equation
\begin{equation}
    \MMB{\uH}{\uW} = 0_4.
\end{equation}
Note that we have used this notation to distinguish this bracket from the Poisson or Moyal bracket.
Algebraically, this is not even a Lie bracket nor a derivation  but we consider it for reasons that will become clear when we discuss deformation quantisation (as might be expected from the importance of Weyl ordering in the Wigner function formulation of phase space quantum mechanics). 

\subsection{Free particle example: Liouville equation}

Let us now evaluate the relativistic Liouville equation~(\ref{MLE} for the free particle). As $\uH_\mathrm{free}(x,p)$ in this example has no explicit $x$ dependence and it is also linear in momentum we have:
\begin{equation}
    \PD{\uH_\mathrm{free}(x,p)}{x^\nu}
    =
    \mathrm{0} 
    \text{ and }
    \PD{\uH_\mathrm{free}(x,p)}{p_\nu}
    =
    c \gamma^0\gamma^\nu
\end{equation}
which leads to
\begin{equation} \label{eq:free1}
\boxed{
    \frac{1}{c}\PD{\uW}{t}
    +\alpha^i\PD{\uW}{x^i}
    =0_4.
    }
\end{equation}
The appearance of $\alpha^i$ in the unprojected equation should not be interpreted as the classical characteristic velocity. 
It rather a kind of ``velocity matrix'' arising from the linear Clifford factor in $\uH_\mathrm{free}$. One would ususaly expect this coefficient to be $p^i/(\gamma m)$ (the classical transport velocity). 
We will now show that is recovered by projection onto the appropriate mass-shell energy sector providing a reassuring link between the ideas proposed in this manuscript and established results. 
Following~\cite{thaller1992dirac} we introduce projection operators
\begin{equation}
    P_\pm(\mathbf p)
    =
    \frac{1}{2}\left(1_4\pm
    \frac{1}{E_p}\uH_\mathrm{3+1}.
    \right)
\end{equation}
Then $P_\pm$ is an idempotent and orthogonal satisfying
\begin{equation}
 P_\pm^2 = P_\pm, \quad P_+P_-=0=P_-P_+, \quad P_++P_-=1_4  
\end{equation}
and, to simplify presentation we will for the moment write\footnote{
    Note that, in general, the decomposition would be
    \begin{align*}
    \uW
    &=(P_+ +P_-)\uW (P_+ +P_-)
    \\
    &=
    P_+\uW P_+
    +
    P_+\uW P_-
    +
    P_-\uW P_+
    +
    P_-\uW P_-.
\end{align*}
So that the diagonal terms carry positive and negative energy populations. The off-diagonal terms encode possible cross-sector correlations which are assume to be zero for this example. The case where this is not the case will be covered in section~\ref{sec:decomp}}
\begin{equation}
    \uW = f_+P_+ + f_-P_-
\end{equation}
Now we consider  the projected equation~(\ref{eq:free1}) which is 
\begin{equation}
    P_\pm\left(\frac{1}{c}\PD \uW t 
    +
    \alpha^i\PD \uW {x^i} \right)P_\pm
    =
    0_4
\end{equation}
Substituting the decomposition of $\uW$ gives
$$
    P_s \left[
        \frac{1}{c}
        \sum_{r=\pm1}
        \PD {f_r}{t}P_r
    +
        \alpha^i
        \sum_{r=\pm1}
        \PD {f_r}{x^i} P_r
    \right] P_s 
    = 0_4.
$$
Distributing the projectors leads to
$$
    \frac{1}{c}\sum_{r=\pm1}
    \PD {f_r}{t} P_s P_r P_s
    +
    \sum_{r=\pm1}
    \PD {f_r}{x^i} P_s \alpha^i P_r P_s
    = 
    0_4.
$$
Using the orthogonality and idempotence relations 
$
    P_s P_r 
    = 
    \delta_{sr} P_s,
$ 
we have 
$
    P_s P_r P_s 
    = 
    \delta_{sr} P_s,
$ 
and
$$
    P_s\alpha^i P_r P_s
    =
    P_s\alpha^i( P_r P_s )
    =
    \delta_{sr}P_s\alpha^iP_s.
$$
Therefore the projected equation reduces to
$$
    \frac{1}{c} \PD{f_s}{t} P_s
    +
    \PD {f_s}{x^i} P_s\alpha^iP_s
    = 
    0_4.
$$
Using the projected velocity identity (see Appendix~\ref{sec:derive_PaP})
$$
    P_s\alpha^iP_s
    =
    s\frac{cp^i}{E_p}P_s,
$$
we obtain
$$
    \frac{1}{c}\PD{f_s} t P_s
    +
    s\frac{cp^i}{E_p}\PD{f_s}{x^i} P_s
    =
    0_4.
$$
Factoring out $P_s$,
$$
    \left( \frac{1}{c}\PD{f_s} t
    +
    s\frac{cp^i}{E_p}\PD{f_s}{x^i}
    \right)
    P_s
    = 0_4.
$$
Since $P_s$ is a non-zero projector, the scalar coefficient must vanish which yields
\begin{equation}
    \PD{f_\pm}{t}
    \pm
    \frac{c^2p^i}{E_p}\, \PD{f_\pm}{x^i} 
    =
    0.
\end{equation}
and given that $E_p=\gamma m c^2$ on the mass shell 
\begin{equation}
\boxed{
    \PD{f_\pm}{t}
    \pm
    \frac{p^i}{\gamma m}\, \PD{f_\pm}{x^i} 
    =
    0.
}
\end{equation}
which produces the same prefactor of $\partial/\partial x^i$ as expected when compared with the dynamical equation of motion generated by $\PB{H_\mathrm{free}}{f}=0$ with the scalar Hamiltonian 
\begin{equation}
    H_\mathrm{free}
    =
    \frac{1}{2m}\left(p^\mu p_\mu - m^2c^2 \right).
\end{equation}
which yields
\begin{equation}
    \partial_t W
    \pm
    \frac{p^i}{\gamma m}\, \PD{W}{x^i} 
    = 0.
\end{equation}
This means that this the required factor ${p^i}/{\gamma m}$ arises either from the quadratic nature of $H_\mathrm{free}$ or from the linear $\uH_\mathrm{free}$ by projecting the equation of motion for $\uW$ onto the spaces of positive and negative energy solutions with the important distinction that the latter contains both while in the forma each must be considered separately.

\subsection{Free particle example: $\MMB{\uH_\mathrm{free}}{\uW}=0$ \label{sec:FPWS}}
Here the free particle dynamics would be given by:
\begin{align}
	\frac{1}{c}\PD{\uW}{t} +
	\frac{1}{2} \PD {}{x^i}\COM{\alpha^i}{\uW}_+ 
    =0_4 \label{reftoremarks}
\end{align}
Since \(\alpha^i\) is constant, this may be written as
$$
    \frac{1}{c}\PD \uW t
    +
    \frac{1}{2}\COM{\alpha^i}{\PD \uW {x^i}}_+
    =
    0_4.
$$
Projecting with $P_s$, we obtain
$$
    P_s
    \left(
        \frac{1}{c}\PD \uW t
    +
        \frac{1}{2}\COM{\alpha^i}{\PD \uW {x^i}}_+
    \right)
    P_s
    =
    0_4.
$$
The first term is treated as in the previous case and
the Weyl-ordered spatial term is
$$
    P_s\COM{\alpha^i}{\PD \uW {x^i}}_+P_s
    =
    P_s
    \left(
        \alpha^i\PD \uW {x^i} +\PD \uW {x^i} \alpha^i
    \right)
    P_s.
$$
Substituting $\uW = f_+ P_+ + f_- P_- $
$$
    \PD \uW {x^i}
    =
    \sum_{r=\pm1}\PD {f_r} {x^i}   P_r,
$$
and factoring $\partial {f_r} / \partial {x^i} $ yields
$$
    P_s\COM{\alpha^i}{\PD \uW {x^i}}_+P_s
    =
    \sum_{r=\pm 1}
    \PD {f_r} {x^i}  
    P_s
    \left(
        \alpha^iP_r+P_r\alpha^i
    \right)
    P_s.
$$
Using the orthogonality of the projectors only the $r=s$ term survives (as 
$
    P_s\alpha^iP_rP_s
    =
    P_s\alpha^i(P_rP_s)
    =
    \delta_{rs}P_s\alpha^iP_s
$) so that 
$$
    P_s[\alpha^i,\partial_iW]_+P_s
    =
    (\partial_i f_s)
    \left(
        P_s\alpha^iP_s+P_s\alpha^iP_s
    \right).
$$
Consequently
$$
    P_s[\alpha^i,\partial_iW]_+P_s
    =
    2\PD {f_s}{x^i}P_s\alpha^iP_s.
$$
The factor $1/2$ in the Weyl-ordered equation cancels this factor of 2, so the projected equation leads once more to the same outcome, namley: 
$$
\boxed{    
    \PD{f_s} t
    +
    s\frac{c^2p^i}{E_p}
    \PD {f_s}{x^i}
    =
    0.
}
$$
Therefore the free Weyl-ordered equation gives the same projected branch transport equations as the projected non-Weyl-ordered free equation. meaning that at this stage both are candidate models. We will return to a discussion about which may be more physical later when we consider the dynamics of a charged particle in a electromagnetic field. Following on the projection theme we next turn to a covariant analysis of the free particle.
\subsection{On the energy-sector decomposition of $\uW$ \label{sec:decomp}}
In general $\uW$ can be decomposed into four quadrants
\begin{equation}
    \uW
    =
    P_+\uW P_+
    +
    P_+\uW P_-
    +
    P_-\uW P_+
    +
    P_-\uW P_-
    \label{eq:general-sector-W}
\end{equation}
which are the components of $\uW$ obtained by projection onto the appropriate energy sectors.
In the non-covariant \(3+1\) formulation, the positive- and
negative-energy parts are
\begin{equation}
   \uW_{++}
   \equiv P_+\uW P_+,
   \mathrm{\ and\ }
   \uW_{--}\equiv P_-\uW P_-,
\end{equation}
while the off-diagonal inter-sector blocks are
\begin{equation}
   \uW_{+-}\equiv P_+\uW P_-,
   \mathrm{\ and\ }
   \uW_{-+}\equiv P_-\uW P_+.
\end{equation}
In the standard Dirac representation,
\begin{equation}
   \alpha^i
   =
   \begin{pmatrix}
       0 & \sigma^i\\
       \sigma^i & 0
   \end{pmatrix},
   \qquad
   \beta
   =
   \begin{pmatrix}
        1_2 & 0\\
       0 & - 1_2
   \end{pmatrix},
\end{equation}
ans the matrix form of the projectors will be
\begin{equation}
   P_\pm
   =
   \frac{1}{2}
   \begin{pmatrix}
       \left(
           1\pm \dfrac{mc^2}{E_p}
       \right) 1_2
       &
       \pm \dfrac{c\,\Vec{\sigma}\cdot\Vec{p}}{E_p}
       \\[1.25em]
       \pm \dfrac{c\,\Vec{\sigma}\cdot\Vec{p}}{E_p}
       &
       \left(
           1\mp \dfrac{mc^2}{E_p}
       \right)  1_2
   \end{pmatrix},
   \label{eq:P-plus-explicit}
\end{equation}
Note that the positive- and negative- energy projectors are not generally respectivly
\begin{equation}
   \begin{pmatrix}
        1_2 & 0\\
       0 & 0
   \end{pmatrix} \text{ and }
   \begin{pmatrix}
       0 & 0\\
       0 &  1_2
   \end{pmatrix}.
\end{equation}
The upper and lower two spinor components are mixed whenever
$p^i \neq 0$. Therefore care needs to be taken identifying with the positive-energy and negative sectors from the projections $P_+\uW P_+$ and $P_-\uW P_-$ respectively.

The one important exception is the rest-frame limit as
$p^i=0$. Then $E_p = mc^2$ and the projectors reduce to
\begin{equation}
   P_+
   =
   \begin{pmatrix}
       \mathbf 1_2 & 0\\
       0 & 0
   \end{pmatrix}
   \text{ and }
   P_-
   =
   \begin{pmatrix}
       0 & 0\\
       0 & \mathbf 1_2
   \end{pmatrix}.
\end{equation}
So in the rest-frame case we can identify the upper-left $2 \times 2$
block with the positive-energy sector and the lower-right $2 \times 2$ block
with the negative-energy sector.

There is another case where this distinction can be made clear which is where either $f_+$ or $f_-$ are zero. 
To see this let us not proceed to the full decomposition of the Liouville equation.
We previously wrote 
\begin{equation}
    \uW_{ss}=f_sP_s,
\end{equation}
and we will now adopt for the off-diagonal cross sector `coherence' blocks the notation 
\begin{equation}
    C_s:=P_s\uW P_{-s}.
\end{equation}
so $\uW$ decomposes into positive and negative energy and coherence terms according to
\begin{equation}
    \uW
    =
    f_+P_+
    +
    f_-P_-
    +
    C_+
    +
    C_- .
    \label{eq:free-block-decomposition}
\end{equation}
The terms \(C_+\) and \(C_-\) are not part of the diagonal energy populations.  
They encode cross-sector coupling (which can be thought as analogous to a coherence structure) between the positive- and negative-energy sectors.

We first consider the ordinary free-particle case and, in order to make the presentation less cluttered we will define a Liouvillian function $\mathcal L(\uW)=\PB{\uH}{\uW}$ (where the Hamiltonian is implied by context and a subscript) which evaluates to
\begin{equation}
    \mathcal L_\mathrm{free}(\uW)
    \equiv
    \frac{1}{c}\frac{\partial \uW}{\partial t}
    +
    \alpha^i\frac{\partial \uW}{\partial x^i}.
    \label{eq:free-ordinary-L}
\end{equation}
The equation of motion is
\begin{equation}
    \mathcal L_\mathrm{free}(\uW)=0.
\end{equation}
Projecting this equation with $P_s$ on both sides
\begin{equation}
    P_s\mathcal L_\mathrm{free}(\uW)P_s
    =
    0.
\end{equation}
Using Eq.~\eqref{eq:free-block-decomposition}, this becomes
\begin{align}
    P_s\mathcal L_\mathrm{free}^{(0)}(f_sP_s)P_s
    &+
    P_s\mathcal L_\mathrm{free}^{(0)}(f_{-s}P_{-s})P_s
     \nonumber \\
    & +P_s\mathcal L_\mathrm{free}^{(0)}(C_+ + C_-)P_s .
    = 0
\end{align}
For the free particle the projectors depend only on momentum and not on spacetime position so there are no projector-derivative terms in this case.
The first two terms we have already considered but the last term adds an additional component to the diagonal projected equation which becomes
\begin{equation}
    0
    =
    \left(
        \frac{1}{c}\frac{\partial f_s}{\partial t}
        +
        s\frac{cp^i}{E_p}\frac{\partial f_s}{\partial x^i}
    \right)P_s
    +
    Q_{ss},
    \label{eq:free-ordinary-diagonal-with-Q}
\end{equation}
where
\begin{equation}
    Q_{ss}
    \equiv
    P_s\mathcal L_\mathrm{free}(C_+ + C_-)P_s .
\end{equation}
More explicitly, because of the ordering in Eq.~\eqref{eq:free-ordinary-L},
\begin{equation}
    Q^\mathrm{free}_{ss}
    =
    P_s\alpha^i
    \frac{\partial C_{-s}}{\partial x^i}
    P_s .
    \label{eq:free-ordinary-Qss}
\end{equation}
Thus \(Q^\mathrm{free}_{ss}\) is not part of \(P_s\uW P_s\).  It is the contribution of the
off-diagonal block \(C_{-s}\) to the projected equation of motion for the diagonal block
\(P_s\uW P_s\).

It is helpful to define the off-diagonal transport matrix (so called because $\alpha^i$ is the prefactor to $\partial \uW /\partial x^i$ in the matrix Liouvile equation)
\begin{equation}
    T_s^i
    \equiv
    P_s\alpha^iP_{-s}.
    \label{eq:free-Gamma-P}
\end{equation}
These are somewhat complex but it is worth noting that in the \emph{rest frame} they simplify to
\begin{equation}
   T_+^i(0)
   =
   \begin{pmatrix}
       0 & \sigma^i\\
       0 & 0
   \end{pmatrix}
   \text{ and }
   T_-^i(0)
   =
   \begin{pmatrix}
       0 & 0\\
       \sigma^i & 0
   \end{pmatrix}.
\end{equation}
The $T_s^i$ are not however the Lorenz boosted version of these matrices - that result will have to wait until we look at the covariant Louiville equation and its projectors later.

Since \(C_{-s}=P_{-s}\uW P_s\), Eq.~\eqref{eq:free-ordinary-Qss} may be written as
\begin{equation}
    Q^\mathrm{free}_{ss}
    =
    T_s^i
    \frac{\partial C_{-s}}{\partial x^i}.
\end{equation}
This term vanishes if the off-diagonal block \(C_{-s}\) is set to zero, but it does not
vanish as an algebraic identity.

Now consider the Weyl-ordered free-particle case and define a Liouvillian function $\mathcal L^W(\uW)=\frac{1}{2}\MMB{\uH}{\uW}$ (where the Hamiltonian is implied by context and a subscript) which evaluates to
\begin{equation}
    \mathcal L_\mathrm{free}^{W}(\uW)
    =
    \frac{1}{c}\frac{\partial \uW}{\partial t}
    +
    \frac{1}{2}
    \left[
        \alpha^i,
        \frac{\partial \uW}{\partial x^i}
    \right]_+ .
\end{equation}
The diagonal projected Weyl equation is
\begin{equation}
    P_s\mathcal L_\mathrm{free}^{W}(\uW)P_s 
    =
    0.
\end{equation}
Using the same block decomposition gives
\begin{equation}
    \left(
        \frac{1}{c}\frac{\partial f_s}{\partial t}
        +
        s\frac{cp^i}{E_p}\frac{\partial f_s}{\partial x^i}
    \right)P_s
    +
    Q^{\rm W}_{ss}
    =
    0,
    \label{eq:free-Weyl-diagonal-with-Q}
\end{equation}
where
\begin{equation}
    Q^{\rm W}_{ss}
    =
    P_s\mathcal \mathcal L_\mathrm{free}^{W}(C_+ + C_-)P_s .
\end{equation}
In terms of the off-diagonal blocks this is
\begin{equation}
    Q^{\rm W}_{ss}
    =
    \frac{1}{2}
    \left(
        T_s^i
        \frac{\partial C_{-s}}{\partial x^i}
        +
        \frac{\partial C_s}{\partial x^i}
        T_{-s}^i
    \right).
    \label{eq:free-Weyl-Qss}
\end{equation}
The Weyl-ordered form therefore treats the two off-diagonal blocks symmetrically and has the same rest-frame limit.

Next we note the transport equations for off-diagonal projections. In the ordinary ordered equation we obtain
\begin{equation}
    P_s\mathcal L_\mathrm{free}^{(0)}(\uW)P_{-s}
    =
    \frac{1}{c}\frac{\partial C_s}{\partial t}
    +
    s\frac{cp^i}{E_p}
    \frac{\partial C_s}{\partial x^i}
    +
    \frac{\partial f_{-s}}{\partial x^i}
    T_s^i 
    = 0.
\end{equation}
For the Weyl-ordered equation 
\begin{equation}
    P_s\mathcal L_{\rm W}^{(0)}(\uW)P_{-s}
    =
    \frac{1}{c}\frac{\partial C_s}{\partial t}
    +
    \frac{1}{2}
    \frac{\partial}{\partial x^i}
    \left(
        f_s+f_{-s}
    \right)
    T_s^i 
    =
    0.
\end{equation}

If $\uW$ is interpreted as a positive spinor-matrix probability density, then a
vanishing population in one branch implies the absence of coherence with that
branch. Consider a positive block matrix
\begin{equation}
    M
    =
    \begin{pmatrix}
        A & B \\
        B^\dagger & D
    \end{pmatrix}
    \geq 0 ,
\end{equation}
one has, construct an arbitrary 4-vector $\Vec w = (\Vec u, \Vec v)$ from two two-vectors $\Vec u$ and $\Vec v$. From $\Vec w ^\dagger M \Vec w$ we find a version of the Cauchy–Schwarz inequality 
\begin{equation}
    \left|\Vec u^\dagger B \Vec v\right|^2
    \leq
    \left(\Vec u^\dagger A \Vec u\right)
    \left(\Vec v^\dagger D \Vec v\right).
    \label{eq:B_CS}
\end{equation}
If $D=0$  the right-hand side of Eq.~\eqref{eq:B_CS} vanishes for all $\Vec u$ and $\Vec v$ so that
$\Vec u^\dagger B \Vec v = 0$ also for all and therefore $B = 0$ 

Applying this result we see that if $f_-=0$ then $C_\pm=0$
provided $\uW$ is positive as it will be for all classical distributions (this would not be guaranteed for the Wigner function which may take on negative values).
It follows that a strictly positive-energy free-particle ensemble has the form $\uW=f_+P_+$ and the projected equation then reduces to
\begin{equation}
    \boxed{
    \partial_t f_+
    +
     \frac{c^2p^i}{E_p}\partial_i f_+ 
    = 0
    }
\end{equation}
and under these conditions of positive $\uW$ and $f_-=0$  we recover exactly the ususal classical dynamics (and we could have done the same for negative energy solutions by setting $f_+=0$) to obtain
\begin{equation}
    \boxed{
    \partial_t f_-
    -
     \frac{c^2p^i}{E_p}\partial_i f_- 
    = 0
    }
\end{equation}

\subsection{Covariant Free particle example: Liouville equation \& $\MMB{\uH_\mathrm{free}^\mathrm{cov}}{\uW}=0$ \label{sec:cov}}
Staring with the ususal Liouville equation and using $\mathcal \uH_\mathrm{free}^\mathrm{cov}$ as defined in~\ref{eq:HFreeCov}
$$
    \PB{ \uH_\mathrm{free}^\mathrm{cov}}{\uW}
    =
    \frac{\partial \mathcal \uH_\mathrm{free}^\mathrm{cov}}{\partial p_\mu}
    \frac{\partial \uW}{\partial x^\mu}
    =
    c\gamma^\mu\PD{\uW}{x^\mu}
    =
    0_4.
$$
which is equivalent to
\begin{equation} \label{eq:covDynamics}
    \gamma^\mu\PD{\uW}{x^\mu}
    =
    0_4.    
\end{equation}
As before let us perform projection of the free-particle equation using but this time we will need to use
the covariant mass-shell idempotents which, following~\cite{BjorkenDrell1964RQM}, are
\begin{equation}
    \Pi_s(p)
    =
    \frac{1}{2}
    \left(
        1_4+s\frac{\gamma^\mu p_\mu}{mc}
    \right),
\end{equation}
where $s=\pm 1$ (note~\cite{BjorkenDrell1964RQM} used the notation $\Lambda_\pm(p)$ which we have changed to avoid confusion with Lorentz transformations).
Just as with $P_\pm$ they obey
$$
    \Pi_s^2         = \Pi_s,\ 
    \Pi_s\Pi_r      = \delta_{sr}\Pi_s, \  
    \Pi_+ + \Pi_-   = 1_4
$$
when $p^\mu p_\mu=m^2c^2$. The projectors $\Pi_\pm$ are idempotents in the same sense as the projectors $P_\pm$.  
The difference is that $\Pi_\pm$ are written covariantly on the mass shell, whereas $P_\pm$ are the corresponding coordinate-time, In other words, the $P_\pm$ are the Hamiltonian, energy-branch version of the covariant $\Pi_\pm$ projectors.

As before we will assume a branch-diagonal decomposition
$$
    \uW = \sum_{r=\pm 1} g_r \Pi_r.
$$
We use the notation $g$ instead of $f$ simply to discriminate between the covariant and non-covariant discussions.
For a free particle, the covariant projectors (as with the energy-branch projectors) depend only on momentum, not on spacetime position so that
$$
    \PD{\uW}{x^\mu}
    =
    \sum_{r=\pm1} \PD {g_r}{x^\mu} \Pi_r.
$$
Now projecting eq~\ref{eq:covDynamics} with $\Pi_s$ 
and substituting the decomposition of $\uW$, we obtain
$$
    \Pi_s\gamma^\mu
    \sum_{r=\pm1}
    \PD{g_r}{x^\mu}\Pi_r\Pi_s
    =
    0_4.
$$
Using orthogonality of the projectors only the $r=s$ term survives:
$$
    \PD{ g_s}{x^\mu}\Pi_s\gamma^\mu\Pi_s
    =
    0_4.
$$
In direct analogy with the $P_s a^i P_s$ we have, with a somewhat more elegant derivation in Appendix~\ref{sec:DerivationPigammaPi}, the covariant equivalent
$$
    \Pi_s\gamma^\mu\Pi_s
    =
    s \frac{p^\mu}{mc}\Pi_s.
$$
Substitution gives
$$
    s\frac{p^\mu}{mc}
    \PD {g_s}{x^\mu}\Pi_s
    =
    0_4.
$$
Since $s/(mc)\neq0$, and since $\Lambda_s$ is a non-zero projector, the scalar transport equation is the covariant branch-transport equation
$$
\boxed{
        p^\mu\PD{g_s}{x^\mu} 
        = 
        0.
      }
$$
We recover the coordinate-time equations by writing
$$
    x^0 = ct 
        \mathrm{\ and\ }
    \PD{}{x^0} = \frac{1}{c}\PD{}{t}.
$$
On the positive-energy sheet,
$$
    p^\mu
    =
    \left(\frac{E_p}{c}, p^i\right),
$$
so
$$
    \boxed{
        \PD {g_+}{t}
    +
        \frac{c^2p^i}{E_p}\PD{g_+}{x^i}
    =
        0.
    }
$$
And on the negative-energy sheet,
$$
    p^\mu
    =
    \left(-\frac{E_p}{c}, p^i\right),
$$
so
$$
    \boxed{
        \PD {g_-}{t}
        -
        \frac{c^2p^i}{E_p}\PD {g_-}{x^i}
        =
        0.
    }
$$
Projection with $\Pi_\pm$ recovers the correct relativistic
branch velocities and does so in covariant form.
 
The proof that the Weyl-Symmetrised bracket $\MMB{\uH_\mathrm{free}^\mathrm{cov}}{\uW}=0_4$ also leads to this same answer is again the deviation of section \label{sec:FPWS} but adapted using the same logic as already used in this section. We have omitted including it for brevity.

In direct analogy with the $3+1$ energy-sector decomposition, the full
spinor-matrix distribution may be decomposed covariantly as
$$
        \uW
        =
        \Pi_+\uW\Pi_+
        +
        \Pi_+\uW\Pi_-
        +
        \Pi_-\uW\Pi_+
        +
        \Pi_-\uW\Pi_- .
$$
We write the diagonal branch populations as
$$
        \uW_{ss}^\Pi
        =
        \Pi_s\uW\Pi_s
        =
        g_s\Pi_s ,
$$
and the off-diagonal cross-sector blocks as
$$
        C_s^\Pi
        =
        \Pi_s\uW\Pi_{-s}.
$$
so
$$
        \uW
        =
        g_+\Pi_+
        +
        g_-\Pi_-
        +
        C_+^\Pi
        +
        C_-^\Pi .
$$
The superscript $\Pi$ is included here only to distinguish these covariant blocks from the earlier $3+1$ blocks $C_s=P_s\uW P_{-s}$.
Since the free-particle projectors depend on momentum but not on spacetime position,
$
        \partial_\mu \Pi_s(p)=0 ,
$
for the free-particle spacetime derivative. So that we again have 
$$
        \PD { \uW}{x^\mu}
        =
        \PD {g_+}{x^\mu}\Pi_+
        +
        \PD {g_-}{x^\mu}\Pi_-
        +
        \PD { C_+^\Pi} {x^\mu}
        +
        \PD { C_-^\Pi} {x^\mu} .
$$
We define the covariant off-diagonal version of the gamma block
$$
        \Gamma_s^\mu(p)
        =
        \Pi_s(p)\gamma^\mu\Pi_{-s}(p)
$$
which is the covariant replacement of the non-covariant
off-diagonal velocity matrix
$$
        \Gamma_s^i=P_s\alpha^iP_{-s}.
$$
Unlike the $3+1$ object, $\Gamma_s^\mu$ transforms with $\mu$ as a spacetime vector index, while each component
acts as a matrix on the spinor-sector indices so that
$$
        \Gamma_s^{\prime\mu}(p')
        =
        \Lambda^\mu{}_\nu\,
        S(\Lambda)\Gamma_s^\nu(p)S^{-1}(\Lambda),
$$
provided that
$
        \Pi'_s(p')
        =
        S(\Lambda) \Pi_s(p) S^{-1} (\Lambda)
$ 
and also that
$
        S^{-1}(\Lambda)\gamma^\mu S(\Lambda)
        =
        \Lambda^\mu{}_\nu\gamma^\nu.
$.

In the positive-energy rest frame where $p^\mu=(mc,\Vec{0})$, and again using the standard Dirac representation, we see
$$
        \Pi_+(0)
        =
        \begin{pmatrix}
        1_2 & 0\\
        0 & 0
        \end{pmatrix}
        \text{ and }
        \Pi_-(0)
        =
        \begin{pmatrix}
        0 & 0\\
        0 & 1_2
        \end{pmatrix},
$$
and hence
$$
        \Gamma_\pm^0(0)=0_4,
        \ 
        \Gamma_+^i(0)
        =
        \begin{pmatrix}
            0 & \sigma^i\\
            0 & 0
        \end{pmatrix} 
        \text{ and }
        \Gamma_-^i(0)
        =
        \begin{pmatrix}
        0 & 0\\
        -\sigma^i & 0
        \end{pmatrix}.
$$
Unlike $P_s\alpha^iP_{-s}$ are $\Gamma _s^\mu$ covariant and transform with both the spinor similarity transformation and the Lorentz vector index. So for example non-rest frame versions can be generated by Lorentz boosts which adds clarity to there role and function.

Now let us reconsider the full diagonal projected equation.
The ordinary covariant free-particle equation is
$$
        \mathcal L_{\mathrm{free}}^{\mathrm{cov}}(\uW)
        =
        \gamma^\mu \PD{\uW}{x^\mu}
        =
        0_4 .
$$
Projecting with $\Pi_s$ on both sides gives
$$
        \Pi_s\mathcal L_{\mathrm{free}}^{\mathrm{cov}}(\uW)\Pi_s 
        =
        0_4.
$$
Using the block decomposition,
$$
\begin{aligned}
        \Pi_s\mathcal L_{\mathrm{free}}^{\mathrm{cov}}(g_s\Pi_s)\Pi_s
        &+
        \Pi_s\mathcal L_{\mathrm{free}}^{\mathrm{cov}}(g_{-s}\Pi_{-s})\Pi_s  \\
        &+
        \Pi_s\mathcal L_{\mathrm{free}}^{\mathrm{cov}}
        (C_+^\Pi+C_-^\Pi)\Pi_s 
        =
        0_4.
\end{aligned}
$$
The first term gives
$$
        \Pi_s\gamma^\mu\PD{ g_s}{x^\mu}\Pi_s
        =
        s\,\frac{p^\mu}{mc}\,\PD{ g_s}{x^\mu}\,\Pi_s ,
$$
while the second diagonal-population term vanishes by orthogonality.
Thus
$$
        s\,\frac{p^\mu}{mc}\,\PD {g_s}{x^\mu}\,\Pi_s
        +
        \mathcal Q^\mathrm{free}_{ss}
        =
        0_4
$$
where
$$
        \mathcal Q^\mathrm{free}_{ss}
        =
        \Pi_s\gamma^\mu\partial_\mu
        (C_+^\Pi+C_-^\Pi)\Pi_s .
$$
from the orthogonality of the projectors only the right hand sided term survives yielding
$$
        \mathcal Q^\mathrm{free}_{ss}
        =
        \Gamma_s^\mu\,\partial_\mu C_{-s}^\Pi .
$$
And so the diagonal branch equation is
$$
\boxed{
        s\,\frac{p^\mu}{mc} \PD{g_s}{x^\mu} \Pi_s
        +
        \Gamma_s^\mu\PD{ C_{-s}^\Pi}{x^\mu} 
        =
        0_4.
}
$$
Now the off-diagonal elements arise from
$$
        \Pi_s\mathcal L_{\mathrm{free}}^{\mathrm{cov}}(\uW)\Pi_{-s}
        =
        0_4
$$
Using the block decomposition gives
$$
        s\,\frac{p^\mu}{mc}\PD{C_s^\Pi}{x^\mu}
        +
        \PD {g_{-s}}{x^\mu}\Gamma_s^\mu 
        =
        0_4.
$$
Thus, in the ordinary covariant equation, the off-diagonal block $C_s^\Pi$ is transported along the same covariant branch but it is also has a contribution from gradients of the opposite diagonal population.

As with the non-covariant case if $g_+$ or $g_-$ is zero and $\uW$ is positive then the second term vanishes usual scalar covariant branch-transport equation is recovered.
$$
\boxed{
    \frac{p^\mu}{mc} \PD{g_\pm}{x^\mu} = 0
}
$$

For the diagonal Weyl-symmetrised covariant ordering we define
$$
        \mathcal L_{\mathrm{W}}^{\mathrm{cov}}(\uW)
        =
        \frac{1}{2c}
        \MMB{\uH_{\mathrm{free}}^{\mathrm{cov}}}{\uW}
        =
        \frac{1}{2}
        \COM{
        \gamma^\mu
        }{
        \PD{\uW}{x^\mu}
        }_+ 
$$
Projecting onto the diagonal sector gives a very similar derivation to the above except the off-diagonal contribution is now symmetrised between the two
cross-sector blocks:
$$
        \mathcal Q^\mathrm{W}_{ss}
        =
        \Pi_s\mathcal L_{\mathrm{W}}^{\mathrm{cov}}
        (C_+^\Pi+C_-^\Pi)\Pi_s ,
$$
or explicitly
$$
        \mathcal Q^\mathrm{W}_{ss}
        =
        \frac{1}{2}
        \left(
        \Gamma_s^\mu\PD{ C_{-s}^\Pi}{x^\mu}
        +
        \PD {C_s^\Pi}{x^\mu}\Gamma_{-s}^\mu
        \right).
$$
The Weyl-symmetrised diagonal projected equation is therfore
$$
\boxed{
        s\,\frac{p^\mu}{mc}\PD{f_s}{x^\mu}\Pi_s
        +
        \frac{1}{2}
        \left(
        \Gamma_s^\mu \PD{ C_{-s}^\Pi}{x^\mu}
        +
        \PD{ C_s^\Pi}{x^\mu}\Gamma_{-s}^\mu
        \right)
        =
        0.
}
$$
For the Weyl-symmetrised covariant ordering off-diagonal projected equations we obtain
$$
        \Pi_s\mathcal L_{\mathrm{W}}^{\mathrm{cov}}(\uW)\Pi_{-s}
        =
        0
$$
which gives
$$
        \frac{1}{2}
        \left[
        \PD{}{x^\mu}(g_s+g_{-s})
        \right]\Gamma_s^\mu 
        =
        0.
$$
Importnatly the self-transport terms proportional to
$p^\mu\partial_\mu C_s^\Pi$ cancel between the two sides of the anti-commutator. This is the crucial difference between the fully covariant Weyl-symmetrised equation and the $3+1$ Weyl-symmetrised equation.: the two formulations agree in the diagonal scalar branch limit,
but they do not assign identical dynamics to the off-diagonal Clifford-sector structure.

Again, if $g_+$ or $g_-$ is zero and $\uW$ is positive then the second term vanishes usual scalar covariant branch-transport equation is recovered as previously stated for the non-symmetrised case.

\subsection{Comparing the normal and Weyl symmetrised brackets}

The agreement of the covariant and \(3+1\) Weyl-symmetrised equations in the diagonal scalar branch limit is an useful consistency check. We have shown that both formulations recover the usual classical relativistic transport equation after making reasonable assumptions ($g_+$ or $g_-$ is zero and $\uW$ is positive) which is an encouraging demonstration of a correspondence principle (by reproducing the predictions of the exiting theory in the limit where it is known to be good).

The two formulations are however dynamically different in diagonal and in particular off-diagonal blocks. It is this distinction that will allow us to test these models against existing classical theories and each other.
So far the fully covariant Weyl-symmetrised equation is the more natural candidate as 
(1) it does not select a preferred time coordinate and it is built directly from the covariant Clifford-valued mass-shell factor and
(2) it is the form most naturally aligned with Weyl/Wigner deformation quantisation (see later).

\section{The Charged Particle \label{sec:cp}}
We now introduce an external electromagnetic field.  This is hope is that there will arise experimentally testable hypothesis of this model and each ordering of the matrix-valued Liouville.  
In an electromagnetic field the projectors depend on phase-space position, and their derivatives generate additional internal Clifford-algebraic structure.

Note that out of necessity we will recycle much of the notation from previous material reusing symbols such as $\pi$, $\Pi$ and $\Gamma$ with similar meaning but now for charged particle in and electromagnetic field. Which notation is being used at any one time should be clear by context.

We introduce the electromagnetic field by the usual 
\begin{equation}
    \pi_\mu(x,p)
    =
    p_\mu-\frac{q}{c}A_\mu(x),
    \label{eq:pi-def}
\end{equation}
The covariant generator is
\begin{equation}
    \uH_{\rm em}^{\rm cov}
    =
    c\left(
        \gamma^\mu\pi_\mu-mc\, 1_4
    \right).
    \label{eq:H-em-cov}
\end{equation}
The corresponding covariant energy-sector projectors are
\begin{equation}
    \Pi_s(\pi_\mu)
    =
    \frac{1}{2}
    \left(
        \mathbf 1_4
        +
        s\frac{\gamma^\mu\pi_\mu}{\kappa}
    \right),
    \qquad
    s=\pm1,
    \label{eq:Lambda-s-def}
\end{equation}
where the invariant mass (rest) mass is
\begin{equation}
    \kappa \equiv \sqrt{\pi^\mu\pi_\mu}.
\end{equation}
Where, on the mass-shell $\kappa=mc$. 
Note that $\Pi_s(\pi_\mu)$ are 
idempotent where $\kappa$ is real and non-zero and in our case we chose it to be on the mass shell.
The projectors obey the usual relationships 
    $\Pi_s^2=\Pi_s$,
    $\Pi_s\Pi_{r}=\delta_{sr}\Pi_s$
    and
    $\Pi_{+} + \Pi_{-} = \mathbf 1_4$.
The projected gamma identity is now
\begin{equation}
    \Pi_s(\pi_\mu)\gamma^\mu\Pi_s(\pi_\mu)
    =
    s\frac{\pi^\mu}{\kappa}\Pi_s(\pi_\mu) .
    \label{eq:Lambda-gamma-Lambda}
\end{equation}
and the off-diagonal gamma block
\begin{equation}
    \Gamma_s^\mu
    \equiv
    \Pi_s(\pi_\mu)\gamma^\mu\Pi_{-s}(\pi_\mu).
    \label{eq:Gamma-s-def}.
\end{equation}
As usual we will write $\uW$ in terms of positive- and negative-energy branch populations and inter-branch coupling
\begin{equation}
    \uW_{sr}
    =
    \Pi_s(\pi_\mu)\uW\Pi_{r}(\pi_\mu)
    \text{ and }
    \uW=\sum_{s,r=\pm1}\uW_{sr}.
\end{equation}
where $s,r=\pm 1$.
As before and for later comparison with standard Liouvillian scalar transport we once more write the diagonal blocks as
\begin{equation}
    \uW_{ss}=g_s\Pi_s(\pi_\mu),
\end{equation}
and the off-diagonal blocks by
\begin{equation}
    C_s:=\Pi_s(\pi_\mu)\uW\Pi_{-s}(\pi_\mu).
\end{equation}
Thus
\begin{equation}
    \uW
    =
    g_+\Pi_{+}(\pi_\mu)
    +
    g_-\Pi_{-}(\pi_\mu)
    +
    C_+
    +
    C_- .
    \label{eq:W_em_decomp}
\end{equation}
Once more the matrix Liouvillian equation can be wreitten as (for either bracket choice)
\begin{equation}
    \mathcal L_\mathrm{em}(\uW)=0
\end{equation}
denote whichever charged-particle Liouville equation is under consideration.  The projected
equation for the \((s,r)\) block is then
\begin{equation}
    \Pi_s(\pi_\mu)\mathcal L_\mathrm{em}(\uW)\Pi_{r}(\pi_\mu) 
    = 
    0.
    \label{eq:projected_L_diag}
\end{equation}
Using Eq.~\eqref{eq:W_em_decomp}, this becomes
\begin{align}
    &\Pi_s(\pi_\mu)\mathcal L_\mathrm{em}(g_s\Pi_s(\pi_\mu))\Pi_{r}(\pi_\mu)
    \nonumber \\ 
    &+ 
    \Pi_s(\pi_\mu)\mathcal L_\mathrm{em}(g_{-s}\Pi_{-s}(\pi_\mu))\Pi_{r}(\pi_\mu)
    \nonumber \\ 
    &+ 
    \Pi_s(\pi_\mu)\mathcal L_\mathrm{em}(C_+ + C_-)\Pi_{r}(\pi_\mu) 
    =
    0
    \label{eq:block-L-expansion}
\end{align}
This equation is the cleanest way to distinguish the decomposition of $\uW$ from the dynamical coupling between blocks.
Importantly in this case of the charged-particle problem the
projectors are now local phase-space projectors because the four potential has spacetime dependence. Consequently the product rule applies and
$$
    \mathcal L_\mathrm{em}(g_s\Pi_s)
    =
    \mathcal L_\mathrm{em}(g_s)\Pi_s
    +
    g_s\mathcal L_\mathrm{em}(\Pi_s),
$$
The second term is absent only in the free-particle
case where the projectors have no explicit spacetime dependence and the relevant Liouvillian contains no force term acting on their momentum dependence. 
We will see that these projector-derivative terms will produce additional Clifford-sector contributions to the dynamics.
For the diagonal block \(r=s\), define
\begin{equation}
    Q^\mathrm{em}_{ss}
    :=
    \Pi_s(\pi_\mu)\mathcal L_\mathrm{em}(C_+ + C_-)\Pi_s(\pi_\mu) .
    \label{eq:Q-general-def}
\end{equation}

Once more we will consider the the two Liouvillians from each candidate bracket. The Poisson bracket matrix-valued Liouvillian evaluates to
\begin{equation}
    \mathcal L_\mathrm{em}^\mathrm{PB}(\uW)
    =
    \gamma^\mu
    \PD{\uW}{x^\mu}
    +
    \frac{q}{c}
    \PD{A_\mu}{x^\nu}
        \PD{\uW} {p_\nu}
    \gamma^\mu
\end{equation}
and the Weyl-symmetrised version to
\begin{equation}
    \mathcal L_\mathrm{em}^\mathrm{W}(\uW)
    =
    \COM{
            \gamma^\mu
        }{
            \PD{\uW}{x^\mu}
    }_+
    +
    \frac{q}{c}
    \PD{A_\mu}{x^\nu}
    \COM{
        \PD{\uW} {p_\nu}
        }{
        \gamma^\mu
    }_+.
\end{equation}

To simplify the analysis and also to give a concrete example we will assume to the Landau gauge
\begin{equation}
    A_y=Bx,
    \qquad
    A_0=A_x=A_z=0,
\end{equation}
so that the magnetic field points in the \(z\)-direction.  Then the momentum is simply
\begin{equation}
    \pi_y=p_y-\frac{qB}{c}x .
\end{equation}
The ordinary Liouvillian becomes
\begin{equation}
    \mathcal L_\mathrm{em}^\mathrm{PB}(\uW)
    =
    \gamma^\mu
    \PD{\uW}{x^\mu}
    +
    \frac{qB}{c}
    \PD{\uW}{p_x}
    \gamma^y .
\end{equation}
The Weyl-symmetrised Liouvillian becomes
\begin{equation}
    \mathcal L_\mathrm{em}^\mathrm{W}(\uW)
    =
    \COM{
        \gamma^\mu
        }{
        \PD{\uW}{x^\mu}
    }_+
    +
    \frac{qB}{c}
    \COM{
        \PD{\uW}{p_x}
        }{
        \gamma^y
    }_+ .
    \label{eq:L-W-Landau}
\end{equation}
and the relevant projector derivatives will be
\begin{equation}
    \PD{\Pi_s(\pi_\mu)}{x}
    =
    -\frac{qB}{c}\PD{\Pi_s(\pi_\mu)}{\pi_y}
    \text{ and }
    \PD{\Pi_s(\pi_\mu)}{p_x}
    =
    \PD{\Pi_s(\pi_\mu)}{\pi_x}.
\end{equation}

Considering the normal Poisson bracket first, the Liouville equation~\ref{eq:projected_L_diag} in the diagonal projection is
\begin{align}
    \frac{s}{\kappa}
    &\left(
        \pi^\mu\PD{g_s}{x^\mu}
        +
        \frac{qB}{c}\pi^y\PD{g_s}{p_x}
    \right) \Pi_s(\pi_\mu)
    \\ & -
    \frac{s}{\kappa}\frac{qB}{c}
    (g_s-g_{-s})
    \Pi_s(\pi_\mu)S^{xy}\Pi_s(\pi_\mu)
    +
    Q^{\rm ord}_{ss}
    = 0 \nonumber
\end{align}
Note that $S^{xy}$ the Clifford-algebra bivector associated with the xy-plane
\begin{equation}
    S^{xy}
    \equiv
    \frac{1}{2}\COM{\gamma^x}{\gamma^y}.
\end{equation}
So that the dynamics are given by three terms which are, respectivly (1) branch transport (2) a spin-bivector term arising from projector derivatives and (3) the projetcion of a cross sector coupling term into each block.

Interestingly the  Weyl symmetrisation cancels the explicit projector-derivative spin-bivector term (which me might expect in the classical limit and the diaonal projection is simply
\begin{equation}
    \frac{s}{\kappa}
    \left(
        \pi^\mu\PD{g_s}{x^\mu}
        +
        \frac{qB}{c}\pi^y\PD{g_s}{p_x}
    \right)\Pi_s(\pi_\mu)
    +
    Q^{\rm W}_{ss}
    =
    0
\end{equation}
In situations where the $C_\pm$ vanish the Weyl symmetrised case reduces to

\begin{equation}
\boxed{    \pi^\mu\PD{ g_s}{x^\mu}
    +
    \frac{qB}{c}\pi^y\PD{g_s}{p_x}
    =
    0.
    }
\end{equation}
which in 3+1 cooridnate-time
\begin{equation}
\boxed{
    \PD{g_s}{t}
    +
    s\frac{c^2\pi^i}{E_\pi}\PD{g_s}{x^i}
    +
    s qB\frac{c\pi^y}{E_\pi}\PD{g_s}{p_x}
    =
    0.}
\end{equation}
Which is in agreement with the standard relativistic Liouville/Vlasov equation for a charged spinless particle in Landau gauge

\section{A minimal argument for quantum mechanics}

Inspired once more by Dirac, but this time from his argument for the Heisenberg matrix formulation ~\cite{dirac1981principles} adapted for phase space arguments~\cite{QM}.
We start by noticing both the relativistic and non-relativistic Poisson bracket satisfy the definition of a Lie algebra:
\begin{align}
\PB{u}{v} &= - \PB{v}{u}, \\
\PB{a u_1 + b u_2}{v} &=  a\PB{u_1}{v} + b\PB{u_2}{v}, \\
\PB{u_1 u_2}{v} &= \PB{u_1}{v}u_2 + u_1\PB{u_2}{v}, \label{eq:PB:key1} \\
\PB{u}{v_1 v_2} &= \PB{u}{v_1}v_2 + v_1\PB{u}{v_2}, \label{eq:PB:key2} \\
\PB{\PB u v}{w} &+\PB{\PB w u}{v} + \PB{\PB v w}{u}  = 0.
\end{align}
where $a$ and $b$ are just numbers. 
These are the rules of infinitesimal transformations, such as rotation or translation. 

Dirac's argument for commutators was that for any dynamical theory, we would like to retain the structure that the state at time $t$ going to $t+\delta t$ is an infinitesimal transformation. 
That argument begins by considering $\QPB{u_1 u_2}{v_1 v_2}$. Apply equation~\ref{eq:PB:key1} to obtain:
\begin{equation}
\QPB{u_1 u_2}{v_1 v_2} = \QPB{u_1}{v_1 v_2}u_2 + u_1\QPB{u_2}{v_1 v_2},
\end{equation}
and then equation~\ref{eq:PB:key2} to obtain:
\begin{equation}
\QPB{u_1 u_2}{v_1 v_2} = \QPB{u_1 u_2}{v_1}v_2 + v_1\QPB{u_1 u_2}{v_2},
\end{equation}
equating these yields
\begin{equation}
\QPB{u_1}{v_1 v_2}u_2 + u_1\QPB{u_2}{v_1 v_2} = \QPB{u_1 u_2}{v_1}v_2 + v_1\QPB{u_1 u_2}{v_2},
\end{equation}
making use of the other relations and simplifying we find:
\begin{equation} \label{eq:COM_PB}
\QPB{u_1}{v_1}\COM{u_2}{v_2} = \COM{u_1}{v_1}\QPB{u_2}{v_2},
\end{equation}
where $\COM{a}{b}=ab-ba$. In this way, the argument for operators observables and commutation relations was made. 
With some work, one arrives at $\COM{\cdot}{\cdot} = \mathrm{i} \hbar\PB{\cdot}{\cdot}$ (where $\hbar$ is some constant to be determined through experiment) as a scheme for quantisation.
This is an imperfect scheme, as seen from Groenewold's theorem.

Phase-space provides an alternative approach. 
Here we seek to replace the relativistic matrix Poisson Bracket with a quantum counterpart which we will denote $\MB{\cdot}{\cdot}$ after the Moyal Bracket that must form a Lie algebra and also demand that 
\begin{equation}
  \lim_{\hbar \Tends 0} \MB{\cdot}{\cdot} = \PB{\cdot}{\cdot},
\end{equation}
in terms of some yet to be determined constant $\hbar$.
This bracket is thus a continuous deformation of the Poisson bracket.
The classical matrix-valued brackets introduced previously are defined using ordinary matrix multiplication. Because their algebra is non-commutative, those brackets are bilinear and antisymmetric but they do not in general satisfy either the Jacobi identity or the Leibniz rule [equations (\ref{eq:PB:key1}) and (\ref{eq:PB:key2})].

Following standard phase-space logic, we now seek a redefinition of multiplication of spinor-matrix distribution functions, $\mathbf{F}$ and $\mathbf{G}$ in terms of a star product $\mathbf{F} \star \mathbf{G}$. Where  
\begin{align}
	  \MB{\mathbf{F}}{\mathbf{G}} =  \frac{1}{\mathrm{i}\hbar}(\mathbf{F} \star \mathbf{G} - \mathbf{G} \star \mathbf{F} ).
\end{align}
is a commutator in the associative star algebra.  It therefore
satisfies antisymmetry and the Jacobi identity, and it is a derivation
with respect to the star product,
\begin{equation}
    \MB{\uF \star \uG}{\uK}
    =
    \uF \star \MB{\uG}{\uK}
    +
    \MB{\uF}{\uK}\star \uG.    
\end{equation}
It is not, however, generally a derivation with respect to ordinary
pointwise matrix multiplication.

The $1/\mathrm{i}\hbar$ allows the non-relativistic approach to achieve this limit in an elegant way.
Because $\uH$ and $\uW$ are now matrix-valued, the ordering of
factors in the bracket is important.  
The main obstruction appears in the classical limit, where, for
matrix-valued phase-space functions,
$$
\MB{\uF}{\uG}
=
\frac{1}{\ui \hbar}\COM{\uF}{\uG}+\Order{\hbar^0}
$$
so the limit \(\hbar\to0\) is not a smooth deformation of the
ordinary scalar Poisson bracket unless the leading matrix commutator vanishes or is mitigated in some other way.
Thus the star commutator is algebraically well behaved as a Lie
bracket in the associative star algebra.  
The difficulty is therefore the singular classical limit caused by the leading matrix commutator term.

Despite this difficulty let us continue the argument and look to define a new product of phase space functions of the form
\begin{equation}
  \mathbf{F} \star \mathbf{G} = \sum_{n=0}^\infty \hbar^n\,\Pi^n(\mathbf{F},\mathbf{G})
\end{equation}
where $\Pi^n(\mathbf{F},\mathbf{G})$ is some satisfactory function of the phase space distributions $\mathbf{F}$ and $\mathbf{G}$. 
If we set the zeroth-order term in this expression as $\mathbf{F}\mathbf{G}$ then we have
$$\lim_{\hbar \Tends 0} \mathbf{F} \star \mathbf{G} = \mathbf{F}\mathbf{G}
$$ 
and we recover usual multiplication of matrices of functions. 
Inspired by the Poisson bracket and being consistent with the usual Moyal star product but being careful to not change the order of the multiplicand and multiplier suggests the next term should be:
\begin{equation}
  \mathbf{F} \star \mathbf{G} = 
  \mathbf{F}\mathbf{G} + \frac{\mathrm{i} \hbar}{2}
  \left(
 \PD{\mathbf{F}}{p_\mu}\PD{\mathbf{G}}{x^\mu} 
- 
\PD{\mathbf{F}}{x^\mu}\PD{\mathbf{G}}{p_\mu}
  \right)
   + \Order{\hbar^2},
\end{equation}
We will assume that the same solution for $\mathbf{F} \star \mathbf{G}$ as found in textbook treatments of the phase space formulation quantum mechanics which is
\begin{equation}
  \uF\star \uG = \sum_{n=0}^\infty {1 \over n!}\left( {\ui \hbar \over 2}\right)^n\Pi^n(\uF,\uG)
\end{equation}
where, to reduced the notational burden, we write for one canonical pair $(q,p)$ as
\begin{equation}
  \Pi^n(\uF,\uG)= \sum_{k=0}^n(-1)^k {n\choose k} 
  \left[
    \HPD{}{p}{(n-k)}\HPD{}{q}{k}
  \uF \right]
  \left[
    \HPD{}{p}{k}\HPD{}{q}{(n-k)}
  \uG \right]
\end{equation}
which can be more compactly written as
\begin{equation}
  \uF \star \uG = \uF 
  \exp \left[
    \frac{\ui \hbar}{2} 
    \left(
    \overset{\leftarrow}
    {\PD{}{p_\mu}}
    \overset{\rightarrow}
    {\PD{}{x^\mu}}
    - 
    \overset{\leftarrow}
    {\PD{}{x^\mu}}
    \overset{\rightarrow}
    {\PD{}{p_\mu}}
    \right)\right] \uG
\end{equation}
where the arrow's indicate the direction in which the derivative is to act. 
Importantly note that it is through the introduction of the Moyal product and non-locality that bring in the angular momentum units of $\hbar$. 
In this way we argue that the algebra of spin arises solely from the requiring that a relativistic theory containing both mass-shell branches but the dimensions of angular momentum arise from the non-locality of quantum mechanics. 

Unlike the scalar case
\begin{align}
  \MB{\uF}{\uG} 
  &= 
  \frac{1}{\ui \hbar}\left( \uF \star \uG - \uG \star \uF \right)
  \\&\neq  
  \frac{2}{\hbar}
    \uF \sin \left[ 
    \frac{\hbar}{2} 
    \left(    
        \overset{\leftarrow}
        {\PD{}{p_\mu}}
        \overset{\rightarrow}
        {\PD{}{x^\mu}}
    - 
        \overset{\leftarrow}
        {\PD{}{x^\mu}}
        \overset{\rightarrow}
        {\PD{}{p_\mu}}
    \right)\right]  \uG .
\end{align}
as derivatives of $\uF$ and $\uG$ may not commute.
And so we arrive at the first issue as previously described
\begin{align}
	  &\MB{\mathbf{\uH}}{\mathbf{\uW}} =  \frac{1}{\ui  \hbar}
      \COM{\mathbf{\uH}}{\mathbf{\uW}} +
	  \nonumber \\ & 
	  \frac{1}{2}\left[
	  \left(
	  \PD{\uH}{p_\mu}
	  \PD{\uW}{x^\mu}
	  +
	  \PD{\uW}{x^\mu}
	  \PD{\uH}{p_\mu}
	  \right)
	  -
	  \left(
	  \PD{\uH}{x^\mu}
	  \PD{\uW}{p_\mu}
	  +
	  \PD{\uW}{p_\mu}
	  \PD{\uH}{x^\mu}
	  \right)
	  \right]
\nonumber \\ &+\Order{\hbar}.
\end{align}
This reduces to the usual scalar Moyal bracket only in a commuting sector, for example when $\uH$, $\uW$, and their relevant derivatives take values in a common commutative subalgebra, or when one of the objects is proportional to the identity in the matrix indices.  
This is a leading-order obstruction to treating the matrix Moyal bracket as a simple deformation of the proposed classical bracket.  A simultaneous
left- and right-stargenvalue condition,
$
\uH\star \uW=\epsilon \uW=\uW\star \uH,
$ 
defines a restricted stationary sector in which the star-commutator
vanishes.
We note that this would define a special class of stationary solutions for an equation of the same form as the time independent Schr\"odinger equation. For this reason we will consider the stargenvalue equations as well as the matrix Moyal bracket in the following analysis.

The challenge will be to see if there is a resolution to such difficulties and verify the assumption that the higher-order corrections follow the pattern as for the non-relativistic case. Let us assume this will work and press on. The relativistic Moyal equation should be of the form:
\begin{equation}
\MB{\uH}{\uW} = 0.    
\end{equation}
Noting that 
$
\MB{\uH}{\uW} = 0
$
if
\begin{equation}
\uH \star \uW = \epsilon \uW
= \uW \star \uH \label{eq:left-right-stargenvalue}
\end{equation}
which is of the same form as the time independent Schr\"odinger stargenvalue equations of non-relativistic quantum mechanics but in phase spacetime and we noet that both of sides of this equation contain all powers of $\hbar$. Importantly note that setting $\epsilon=0$ will enforce the mass shell for non-trivial solutions of the left and right eigenvalue equation so that
\begin{equation}
\uH \star \uW = 0
= \uW \star \uH \label{eq:left-right-stargenvalue-epsilon-zero}
\end{equation}

\subsection{Free particle}
Only the first-order momentum derivatives do not vanish so we have
\begin{align}
\uH_\mathrm{free} \star \uW
&=
\left(c\gamma^\mu p_\mu-mc^2 \right)  \uW + \frac{\ui \hbar}{2}c\gamma^\nu \PD{\uW}{x^\nu}\\
\uW \star \uH_\mathrm{free}
&=
\uW \left(c\gamma^\mu p_\mu-mc^2 \right) - \frac{\ui \hbar}{2}\PD{\uW}{x^\nu}c\gamma^\nu     
\end{align}
so the stargenvalue equations are
\begin{align}
\left(c\gamma^\mu p_\mu-mc^2 \right)  \uW + \frac{\ui \hbar}{2}c\gamma^\nu \PD{\uW}{x^\nu}
&=\varepsilon\uW
\\
\uW \left(c\gamma^\mu p_\mu-mc^2 \right) - \frac{\ui \hbar}{2}\PD{\uW}{x^\nu}c\gamma^\nu     & = \varepsilon\uW
\end{align}
and the covariant matrix Moyal bracket would be
\begin{equation}
 \MB{\uH_\mathrm{free}}{\uW}  = 
 \frac{c}{\ui \hbar}p_\mu
 \COM{\gamma^\mu}{W}
 +
 \frac{c}{2} \PD{}{x^\mu}\left(\gamma^\mu\uW+\uW\gamma^\mu\right)=0
\end{equation}
or
\begin{equation} 
 p_\mu
 \COM{\gamma^\mu}{W}
 +
 \frac{\ui \hbar}{2}
  \PD{}{x^\mu}\COM{\gamma^\mu}{\uW}_+=0
\end{equation}

\subsection{Charged particle in an electromagnetic field}
Again the covariant Hamiltonian is
\begin{align}
  \uH(x,p) &=
  c\gamma^\mu \pi_\mu(x,p)-mc^2\, 1_{4}  \nonumber \\
  &=\gamma^\mu \left(cp_\mu-q\,A_\mu(x)\right)-mc^2\, 1_{4} \nonumber \\
&= \uH_\mathrm{free} - q\gamma^\mu \,A_\mu(x) \nonumber 
\end{align}
Because $\uH$ is linear in $\Pi_\mu$ the only surviving momentum derivative is:
\begin{align}
  \frac{\partial \uH}{\partial p_\nu}&=c\gamma^\nu.
\end{align}
However the position derivatives will be 
\begin{align}
  \left(\frac{\partial}{\partial x^\nu}\right)^n \uH
    &=-q\,\gamma^\mu\left(\frac{\partial}{\partial x^\nu}\right)^n{ A_\mu}
\end{align}
and will only vanish when $\left(\frac{\partial}{\partial x^\nu}\right)^n{ A_\mu}=0$.
\begin{widetext}
The left stargenvalue equation is:
\begin{align*}
{\uH}\star{\uW} = \varepsilon \uW
={\uH}{\uW} 
+\frac{\ui \hbar}{2}
    \left(
        \PD{\uH}{p_\nu}\PD{\uW}{x^\nu}
        -
        \PD{\uH}{x^\nu}\PD{\uW}{p_\nu}
    \right)
+ \frac{1}{2!}
  \left(
    \frac{i\hbar}{2}
  \right)^{2}
  \left(
        \HPD{\uH}{{x^\nu}}{2}
        \HPD{\uW}{p_\nu}{2}
        - 
        2
        \frac{\partial^2\uH}{\partial x^\nu\partial p_\nu}
        \frac{\partial^2\uW}{\partial x^\nu\partial p_\nu}
        + 
        \HPD{\uH}{{p_\nu}}{2}
        \HPD{\uW}{x^\nu}{2}    \right) \\
    + \frac{1}{3!}\!\left(\frac{i\hbar}{2}\right)^{\!3}
        \left(
        \HPD{\uH}{{p_\nu}}{3}
        \HPD{\uW}{{x^\nu}}{3}    
        -
        3
        \frac{\partial^3\uH}{\partial x^\nu\partial p_\nu^2}
        \frac{\partial^3\uW}{\partial {x^\nu}^2\partial p_\nu}
        +
         3
        \frac{\partial^3\uH}{\partial {x^\nu}^2\partial p_\nu}
        \frac{\partial^3\uW}{\partial x^\nu\partial p_\nu^2}    
          -       
        \HPD{\uH}{{x^\nu}}{3}
        \HPD{\uW}{p_\nu}{3} 
        \right) 
    + \mathcal{O}\left(\hbar^{4}\right)
= \varepsilon \uW
\\
\left[{\gamma^\mu \left(cp_\mu-q\, A_\mu(x)\right)-mc^2 1_{4}}\right]{\uW}+\frac{\ui \hbar}{2}
\left(
q\gamma^\mu\PD{A_\mu}{x^\nu}\PD{\uW}{p_\nu}+c\gamma^\mu\PD{\uW}{x^\nu}\right)
+
\frac{1}{2!}
  \left(
    \frac{i\hbar}{2}
  \right)^{2}
  \left(
  -q\gamma^\mu\frac{\partial^2 A_\mu}{\partial x^\nu\partial x^\rho}
\frac{\partial^2 W}{\partial p_\nu\partial p_\rho}
   \right)  +\mathcal{O}\left(\hbar^{3}\right)
=\varepsilon \uW
\end{align*}

\end{widetext}
we omit the other stargenvalue equation for brevity.
As an aside note that looking for non-trivial solutions to this stargenvalue equation when $\varepsilon=0$ would mean solving
\begin{align*}
&\left[{\gamma^\mu \left(cp_\mu-q\, A_\mu(x)\right)-mc^2 1_{4}}\right]{\uW}
\\
&\ \ \ \ \ \ \ \ \ 
+
    \frac{\ui \hbar}{2}
    \left(
    q\gamma^\mu\PD{A_\mu}{x^\nu}\PD{\uW}{p_\nu}
    + 
    c\gamma^\nu\PD{\uW}{x^\nu}
    \right)
    + 
    \mathcal{O}\left(\hbar^{2}\right)
=0.
\end{align*}
This appears to similar in form to the Dirac equation in phase space seen in the literature~\cite{D1,D2,D3,D4}.

The Moyal bracket will be
\begin{align}
     \MB{\uH}{\uW}
&=
    \frac{1}{\ui \hbar}
    (cp_\mu-qA_\mu) \COM{\gamma^\mu}{\uW}
\nonumber \\ &\quad
    +
    \frac{c}{2}\,
    \PD{}{x^\nu}\COM{\gamma^\nu}{\uW}_{+}
    +
    \frac{q}{2}\,
    \PD{A_\mu}{x^\nu}\,
    \PD{}{p_\nu}\COM{\gamma^\mu}{\uW}_{+}
\nonumber \\ &\quad \nonumber
    -\frac{i\hbar q}{8}\,
    \frac{\partial^2 A_\mu}
        {\partial x^\nu \partial x^\rho} \,
    \frac{\partial^2}{\partial {p_\nu}\partial {p_\rho} }
    \COM{\gamma^\mu}{\uW}
    +O(\hbar^2)\\
    & =0.
\end{align}
so we have
\begin{align}
    &(cp_\mu-qA_\mu) \COM{\gamma^\mu}{\uW}
\nonumber \\ &\quad
    +
    \ui \hbar\frac{c}{2}\,
    \PD{}{x^\nu}\COM{\gamma^\nu}{\uW}_{+}
    +
    \ui \hbar \frac{q}{2}\,
    \PD{A_\mu}{x^\nu}\,
    \PD{}{p_\nu}\COM{\gamma^\mu}{\uW}_{+}
\nonumber \\ &\quad \nonumber
    +\frac{\hbar^2 q}{8}\,
    \frac{\partial^2 A_\mu}
        {\partial x^\nu \partial x^\rho} \,
    \frac{\partial^2}{\partial {p_\nu}\partial {p_\rho} }
    \COM{\gamma^\mu}{\uW}
    +O(\hbar^3)\\
    & =0.
\end{align}
Note that in quantum mechanics higher momentum derivatives of $\uW$ enter the equation. but importantly it remains first order in time. Also note that odd Moyal orders give anticommutator-type derivative terms while even Moyal orders give commutator-type derivative terms.

\subsection{The Landau gauge}
We will set $A_y=Bx$ and $A_0=A_x=A_z=0$ that equates to a constant magnetic field, $B$, in the $z$ direction so that $\pi_y=p_y-qBx/c$. 
The left stargenvalue equation is
\begin{align*}
&\left[{c\gamma^\mu p_\mu-q\gamma^y Bx-mc^2 1_{4}}\right]{\uW}
\\
&\ \ \ \ \ \ \ \ \ +\frac{\ui \hbar}{2}
\left(
qB\gamma^y\PD{\uW}{p_x}+ c\gamma^\nu\PD{\uW}{x^\nu}\right)
=\varepsilon \uW.
\end{align*}
The equation of motion will then be
\begin{align}   
c p_\mu\COM{\gamma^\mu}{\uW}
-
qBx\COM{\gamma^y}{\uW}
+ \quad \quad \quad
& \nonumber  \\
\frac{i\hbar c}{2}\PD{}{x^\nu} \COM{\gamma^\nu}{\uW}_{+}
+ 
\frac{i\hbar qB}{2}\PD{}{p_x}\COM{\gamma^y}{\uW}_{+}
=0 &.
\end{align}

\subsection{Semiclassical interpretation of the matrix Moyal equation}

We have previously flagged a potentials issue with the matrix-valued Moyal bracket considered above. Recal that is has the expansion
\begin{align}
    \MB{&\uH}{\uW}
    =
    \frac{1}{\ui\hbar}
    \COM{\uH}{\uW}
    +
    \nonumber \\ \nonumber 
    & 
    \frac{1}{2}
    \left(
        \PD{\uH}{p_\mu}
        \PD{\uW}{x^\mu}
        +
        \PD{\uW}{x^\mu}
        \PD{\uH}{p_\mu}
        -
        \PD{\uH}{x^\mu}
        \PD{\uW}{p_\mu}
        -
        \PD{\uW}{p_\mu}
        \PD{\uH}{x^\mu}
    \right)
    \\ 
    &
    +
    \Order{\hbar}. \label{eq:restate-moyal-braket}
\end{align}
The commutator does not in general vanish when the elements of the Moyal bracket are matrix valued.
For the usual scalar deformation quantisation of classical phase space this term vanishes and allows for a smooth quantum-to-classical transition when $\hbar \Tends 0$.
The question we seek to answer in this section is whether or not the presence of $\COM{\uH}{\uW}$ negates the validity of the arguments  presented here.

For us to explore the effect of this term in the quantum-to-classical transition let us write the phase space matrix as a power series expansion in Plank's constant 
\begin{equation}
    \uW
    =
    \uW_0
    +
    \hbar \uW_1
    +
    \hbar^2 \uW_2
    +
    \Order{\hbar^3} .    
\end{equation}
and by substituting this expansion into equation~(\ref{eq:restate-moyal-braket}) and collecting powers of $\hbar$ where each must equal zero gives, at order $\hbar^{-1}$,
\begin{equation} \label{eq:zreoth-order-constraint}
\COM{\uH}{\uW_0}=0.
\end{equation}
Which provides the constraint on the zeroth order term that it must commute Hamiltonian. To see if this is reasonable let us reconsider the case of the charged particle in an electromagnetic field.
Recall that the Hamiltonian gnerator is
$$
    H
    =
    c \gamma^\mu\pi_\mu
    -
    mc^2 1_4,
$$
so that in this specific example eq.~\ref{eq:zreoth-order-constraint} simplifies to
\begin{equation}
    \COM{\gamma^\mu\pi_\mu}{\uW_0}
    =
    0.    \label{eq:w0commutator}
\end{equation}
Noting that we can write
   \begin{align*}
    \uW_0
    &=(\Pi_+ + \Pi_-)\uW_0 (\Pi_+ +\Pi_-)
    \\
    &=
    \Pi_+\uW_0 \Pi_+
    +
    \Pi_+\uW_0 \Pi_-
    +
    \Pi_-\uW_0 \Pi_+
    +
    \Pi_-\uW_0 \Pi_-
\end{align*}
Now projecting eq~\ref{eq:w0commutator} yields
\begin{align}
    \Pi_s \COM{\gamma^\mu\pi_\mu}{\uW_0} \Pi_r
    &=
    0 \\
    \Pi_s (\gamma^\mu\pi_\mu \uW_0 -\uW_0 \gamma^\mu\pi_\mu) \Pi_r
    &=
    0    
    \\
    \Pi_s \gamma^\mu\pi_\mu \uW_0 \Pi_r
    -
    \Pi_s \uW_0 \gamma^\mu\pi_\mu \Pi_r
    &=
    0.    \label{eq:w0commutator-last}
\end{align}
Now we note that 
$$
    (\gamma^\mu\pi_\mu)\Pi_s
    =
    \Pi_s(\gamma^\mu\pi_\mu)
    =
    s\kappa\Pi_s
$$
in other words $\Pi_s$ projects onto the eigenspace of
$
    \gamma^\mu\pi_\mu
$
with eigenvalue $s\kappa$. This means that eq~\ref{eq:w0commutator-last} simplifies to
$$
(s-r)\kappa \Pi_s \uW_0 \Pi_r = 0
$$
when $s\neq r$ then $(s-r)\kappa \neq 0$ which means that $\Pi_s \uW_0 \Pi_r = 0$. This means that for $\COM{\uH}{\uW_0}=0$ that $\uW_0$ must be block diagonal, in other words
$$
\uW_0 = \Pi_+ \uW_0 \Pi_+ + \Pi_- \uW_0 \Pi_-
$$
let us define $\uW_0^+=\Pi_+ \uW_0 \Pi_+$ and $\uW_0^-=\Pi_- \uW_0 \Pi_-$.

Now the oder $\hbar^0$ of equation~(\ref{eq:restate-moyal-braket}) will be
\begin{align}
    &\frac{1}{\ui}
    \COM{\uH}{\uW_1}
    +
    \nonumber \\ \nonumber 
    & 
    \frac{1}{2}
    \left(
        \PD{\uH}{p_\mu}
        \PD{\uW_0}{x^\mu}
        +
        \PD{\uW_0}{x^\mu}
        \PD{\uH}{p_\mu}
        -
        \PD{\uH}{x^\mu}
        \PD{\uW_0}{p_\mu}
        -
        \PD{\uW_0}{p_\mu}
        \PD{\uH}{x^\mu}
    \right)
    \\
    &=0
\end{align}
Now we will use the fact that $W_0$ is block diagonal and note that 
$$
\Pi_s \COM{\uH}{\uW_1} \Pi_s =0
$$
because inside a projected energy sector $\uH$  acts as a scalar multiple of the projector because $\gamma^\mu \pi_\mu$ does. This leads to an equation of motion of the form
$$
        \PD{\uH}{p_\mu}
        \PD{\uW_0}{x^\mu}
        +
        \PD{\uW_0}{x^\mu}
        \PD{\uH}{p_\mu}
        -
        \PD{\uH}{x^\mu}
        \PD{\uW_0}{p_\mu}
        -
        \PD{\uW_0}{p_\mu}
        \PD{\uH}{x^\mu}=0
$$
Which is exactly
$$
\boxed{  \MMB{\uH}{\uW_0} = 0_4.}
  $$

This analysis is therefore consistent with our earlier discussion of the covariant matrix-Liouville equation in Sections~\ref{sec:cov} and~\ref{sec:cp}.
the covariant branch projectors we introduced as a natural way of decomposing the spinor-matrix distribution into positive- and negative-energy sectors. The semiclassical expansion of the matrix Moyal equation now gives a complementary justification for this
decomposition. At order $\hbar^{-1}$, the equation
$$
    \COM{\uH}{\uW_0}
    =
    0
$$
implies, away from degeneracies, that the leading semiclassical distribution is block diagonal with respect to the covariant idempotents so that 
$$
    \uW_0
    =
    \Pi_+\uW_0\Pi_+
    +
    \Pi_-\uW_0\Pi_- 
    =
    \uW_0^+ + \uW_0^-.
$$
In this way we see that the leading classical limit suppresses positive-to-negative branch coupling/coherences.
Iterestingly this result does not require the scalar-branch ansatz $\uW_0=g_+\Pi_+ + g_-\Pi_-$.

Since each $\Pi_s$ is a rank-two projector, the most general leading-order branch components $\uW_0^\pm$ may still contain a non-trivial $2\times2$ spin-density structure internal to the corresponding covariant energy sector. 
The scalar form $\uW_s=g_s\Pi_s$ should therefore be considered as an unpolarised or spin-average.

At order $\hbar^0$, the diagonal projection of the commutator term involving $\uW_1$ vanishes within each branch. The remaining projected equation is
therefore a covariant transport equation for the branch density $\uW_s$. This reduces to the same scalar relativistic transport equation obtained in the earlier covariant free-particle and charged-particle analyses. 
In particular, in the Landau-gauge example, the Weyl-symmetrised equation reduces to the usual spinless relativistic Liouville-Vlasov equation for $g_s$.

The semiclassical Moyal analysis supports the structure
developed in the preceding sections. It shows that branch diagonality is not an imposed simplification, but emerges naturally at leading semiclassical order approximation. The scalar branch equations are recovered as a consistent spin-averaged limit, whereas the more general theory contains internal spin-density transport within each covariant energy sector and higher-order inter-branch coherences suppressed by the
positive-negative branch separation. 
In this sense, the framework appears to interpolate between a classical scalar relativistic distribution and
a spinor-matrix phase-space theory.

A full identification with the semiclassical limit of the
Dirac-Wigner theory would require deriving the accompanying constraint equations and showing explicitly how the internal spin-density transport reduces to the known spin-precession dynamics in the appropriate limit. We take a first look at a comparison with the Dirac-Wigner in the next section.

\subsection{Relation to the Dirac--Wigner equation}

Following~\cite{D1,D2,D3,D4} we seek a direct comparison of the matrix valued Moyal equation with the Dirac equation though considering the Wigner transform of the Dirac density matrix.  Let $\psi(x)$ satisfy
the minimally coupled Dirac equation
$$
    \left[
        \ui\hbar c\gamma^\mu
        \left(
            \PD{}{x^\mu}
            +
            \frac{iq}{\hbar c}A_\mu
        \right)
        - mc^2
    \right]
    \psi(x)
    =
    0.
$$
The spinor density is then defiend as
$$
    \rho_{\alpha\beta}(x_1,x_2)
    \equiv
    \psi_\alpha(x_1)\bar\psi_\beta(x_2),
$$
where $\bar \psi = \psi^\dag \gamma^0$ is the Dirac adjoint.
We now introduce the centre and relative coordinates
\[
\xi=\frac{x_1+x_2}{2},
\qquad
y=x_1-x_2.
\]
A spinor-matrix Wigner function is then given by performing a Wigner transformation (i.e. a Fourier transform with respect to the coordinate separation) on $\rho_{\alpha\beta}(x_1,x_2)$, namley
$$
W_{\alpha\beta}(\xi,p)
=
\int \ud^4y\,
\exp\!\left(-\frac{i}{\hbar}p_\mu y^\mu\right)
\rho_{\alpha\beta}\left(\xi+\frac{y}{2},\xi-\frac{y}{2}\right),
$$
with a gauge link inserted between the two spacetime points in the
fully gauge-covariant formulation (see e.g.~\cite{D1}). According to~\cite{wilsonLine}, in a fully gauge-covariant treatment the density should be
connected by a Wilson line.  Our treatment could therefore be regarded as a
fixed-gauge or local-potential version of such treatments.

Applying the Dirac operator to the left coordinate $x_1$ and then
Wigner transforming gives a left star equation
$$
H\star W=0,
$$
where
$$
H(X,p)
=
    \gamma^\mu(c p_\mu - qA_\mu(X))-mc^2 \, 1_4.
$$
Similarly, applying the adjoint Dirac equation to the right coordinate
$x_2$ gives
$$
W\star H=0.
$$
So we find that the Dirac equation for the spinor amplitude becomes a pair of left and right phase-space equations for the spinor-matrix Wigner
function:
$$
H\star W=0,
\qquad
W\star H=0.
$$
which is consistent with the right and left eigenvalue equation eq~(\ref{eq:left-right-stargenvalue-epsilon-zero}) where $\epsilon =0$ (i.e. on the mass-shell). Thus the matrix theory appear to be in agreement with the phase-space Dirac equation. 

To complete the analogy to the previous section, taking the difference of the two Dirac-Wigner equations gives the matrix Moyal-bracket equation
$$
H\star W-W\star H=0,
$$
or $\MB{H}{W}=0$, the usual transport equation.

While these results are encouraging, establishing full equivalence with the Dirac-Wigner theory, and deriving the standard spin-precession dynamics from the internal branch-density equation, remain important tests of the framework. 

\section{Concluding remarks}
This work has explored a route from relativistic statistical mechanics to spinor phase-space structure. 
The starting point was asserting that a relativistic statistical theory should retain both mass-shell branches while remaining first order in phase-spacetime. 
Applying a Clifford factorisation of the relativistic constraint then leads naturally to a $4\times4$ spinor-matrix distribution function. 
Importantly, spinor structure is not introduced as an independent assumption, but arises from the attempt to express relativistic statistical mechanics in a first-order phase-space form that contains both mass-shell branches.

In the free particle case, projection onto the positive- and negative-energy sectors recovers the usual relativistic Liouville transport equations. 
The covariant projectors treatment showed how off-diagonal sector terms may enter when both branches are retained.
For the charged-particle example, the Weyl-symmetrised form reproduces the standard spinless Liouville-Vlasov equation in the diagonal scalar limit, while the additional Clifford-sector terms indicate where spinorial structure enters the theory.

We showed that deformation-quantisation the star product introduces non-commutativity and non-locality in a way that naturally supports spinor degrees of freedom. 
In particular, the left and right stargenvalue equations on the mass-shell
$$
\uH\star \uW = 0 = \uW \star \uH
$$
provide the appropriate phase-space analogue of the Dirac constraint. 
Their commutator part gives the transport equation, while their anticommutator part encodes the mass-shell and spinor constraints. This shows how the present
framework is consistent with the Dirac-Wigner formulation.

The main conclusion is therefore that spinor structure can be understood as emerging naturally from a first-order relativistic statistical theory that retains both branches of the mass shell. 
The requirement of a linear mass-shell factorisation leads to the Dirac-Clifford algebra, while deformation quantisation connects this statistical structure to spinor quantum mechanics in a way that appears to be consistent with the Dirac-Wigner formulation. 
Further work is needed to develop the electromagnetic
case in a fully gauge-covariant form, to analyse the physical interpretation of the off-diagonal sector terms, and to derive the standard spin-precession
limits explicitly. 
Nevertheless, the results suggest a coherent phase-space
route from relativistic statistical mechanics to spinor quantum mechanics

\acknowledgements
I am grateful to those who commented on earlier versions of this work. 
In particular, I am indebted to Tim Spiller for detailed feedback and incisive questions, especially his challenge on how to interpret the classical limit of spin in the absence of Planck's constant. 
I thank John Samson for repeatedly identifying mistakes in those drafts. 
I also thank Todd Tilma, Russell Rundle, and Kieran Bjergstrom for constructive discussions that helped test the novelty of the ideas and clarify their presentation, and Alexander Balanov and Alexandre Zagoskin for additional stimulating conversations. 
Ben Davies kindly suggested improvements that enhanced the readability of the manuscript.

The author also acknowledges the use of successive versions of OpenAI's ChatGPT during the development of this work. 
Over the period in which this problem was studied, the capability of these systems improved substantially,
particularly for checking algebraic consistency, exploring alternative formulations, identifying possible gaps in reasoning, and improving the clarity of the presentation. 
The scientific interpretation, final arguments, and any remaining errors are mine alone\footnote{I offer this work with caution: the central argument may already be known, may be a reformulation of something standard, or may contain an overlooked error. Corrections and references to prior work would be gratefully received.}, and I welcome further corrections or insights from readers.
\appendix

\section{Derivation of $P_s \alpha^i P_s$\label{sec:derive_PaP}}

\emph{Note much of the output for both appendices was generated by ChatGPT and is presented with little modification except making notation consistent and removing some unnecessary lines. ChatGPT uses a key observation that ${\partial \uH_{3+1}}/{\partial p_i}
=
c\alpha^i$ but does not make this clear from the outset.}

The energy-sector projectors are
\[
P_\pm(\mathbf p)
=
\frac{1}{2}
\left(
1_4\pm\frac{\uH_{3+1}(\mathbf p)}{E_p}
\right).
\]
Equivalently, writing \(s=\pm1\), we define
\[
P_s(\mathbf p)
=
\frac{1}{2}
\left(
1_4+s\frac{\uH_{3+1}(\mathbf p)}{E_p}
\right).
\]
The key algebraic identity is
\[
\uH_{3+1}^2=E_p^2\,1_4.
\]
It follows that
\[
\uH_{3+1}P_s=sE_pP_s,
\qquad
P_s \uH_{3+1}=sE_pP_s.
\]
Indeed,
\[
\uH_{3+1}P_s
=
\uH_{3+1}
\frac{1}{2}
\left(
1_4+s\frac{\uH_{3+1}}{E_p}
\right),
\]
so
\[
\uH_{3+1}P_s
=
\frac{1}{2}
\left(
\uH_{3+1}
+
s\frac{\uH_{3+1}^2}{E_p}
\right).
\]
Using \(\uH_{3+1}^2=E_p^2\,1_4\), this becomes
\[
\uH_{3+1}P_s
=
\frac{1}{2}
\left(
\uH_{3+1}
+
sE_p\,1_4
\right).
\]
But
\[
sE_pP_s
=
sE_p
\frac{1}{2}
\left(
1_4+s\frac{\uH_{3+1}}{E_p}
\right)
=
\frac{1}{2}
\left(
sE_p\,1_4+\uH_{3+1}
\right).
\]
Therefore
\[
\uH_{3+1}P_s=sE_pP_s.
\]
The same calculation on the other side gives
\[
P_s \uH_{3+1}=sE_pP_s.
\]

We next derive the projected velocity identity. Differentiate
\[
\uH_{3+1}P_s=sE_pP_s
\]
with respect to \(p_i\). This gives
\[
\frac{\partial \uH_{3+1}}{\partial p_i}P_s
+
\uH_{3+1}\frac{\partial P_s}{\partial p_i}
=
s\frac{\partial E_p}{\partial p_i}P_s
+
sE_p\frac{\partial P_s}{\partial p_i}.
\]
Now premultiply by \(P_s\):
\[
P_s\frac{\partial \uH_{3+1}}{\partial p_i}P_s
+
P_s \uH_{3+1}\frac{\partial P_s}{\partial p_i}
=
s\frac{\partial E_p}{\partial p_i}P_s
+
sE_p P_s\frac{\partial P_s}{\partial p_i}.
\]
Using
\[
P_s \uH_{3+1}=sE_pP_s,
\]
the second term on the left becomes
\[
P_s \uH_{3+1}\frac{\partial P_s}{\partial p_i}
=
sE_pP_s\frac{\partial P_s}{\partial p_i}.
\]
This cancels the final term on the right. Hence
\[
P_s\frac{\partial \uH_{3+1}}{\partial p_i}P_s
=
s\frac{\partial E_p}{\partial p_i}P_s.
\]
But
\[
\frac{\partial \uH_{3+1}}{\partial p_i}
=
c\alpha^i,
\]
and
\[
\frac{\partial E_p}{\partial p_i}
=
\frac{c^2p^i}{E_p}.
\]
Therefore
\[
P_s c\alpha^i P_s
=
s\frac{c^2p^i}{E_p}P_s.
\]
Dividing by \(c\), we obtain
\[
P_s \alpha^i P_s
=
s\frac{cp^i}{E_p}P_s.
\]
Thus the projected Dirac velocity matrix gives the usual classical relativistic velocity on each energy branch.

\section{Derivation of $\Pi_s\gamma^\mu\Pi_s$\label{sec:DerivationPigammaPi}}
$$
    \Pi_s\gamma^\mu\Pi_s
=
    s \frac{p^\mu}{mc}\Pi_s.
$$
This follows as follows.  Let
\[
A=\frac{\gamma^\nu p_\nu}{mc}.
\]
On the mass shell,
\[
p^\mu p_\mu=m^2c^2,
\]
and therefore
\[
A^2=1_4.
\]
The projectors are
\[
\Pi_s=\frac{1}{2}(1_4+sA).
\]
Using the Clifford relation
\[
\{\gamma^\mu,\gamma^\nu\}=2\eta^{\mu\nu}1_4,
\]
we have
\[
A\gamma^\mu+\gamma^\mu A
=
\frac{1}{mc}
\left(
\gamma^\nu p_\nu\gamma^\mu
+
\gamma^\mu\gamma^\nu p_\nu
\right)
=
\frac{2p^\mu}{mc}1_4.
\]
Now compute
\[
\Pi_s\gamma^\mu\Pi_s
=
\frac{1}{4}
(1_4+sA)\gamma^\mu(1_4+sA).
\]
Expanding gives
\[
\Pi_s\gamma^\mu\Pi_s
=
\frac{1}{4}
\left(
\gamma^\mu
+
sA\gamma^\mu
+
s\gamma^\mu A
+
A\gamma^\mu A
\right).
\]
Using
\[
A\gamma^\mu+\gamma^\mu A
=
\frac{2p^\mu}{mc}1_4,
\]
this becomes
\[
\Pi_s\gamma^\mu\Pi_s
=
\frac{1}{4}
\left(
\gamma^\mu
+
s\frac{2p^\mu}{mc}1_4
+
A\gamma^\mu A
\right).
\]
To simplify the final term, multiply
\[
A\gamma^\mu+\gamma^\mu A
=
\frac{2p^\mu}{mc}1_4
\]
on the right by \(A\).  Since \(A^2=1_4\), this gives
\[
A\gamma^\mu A+\gamma^\mu
=
\frac{2p^\mu}{mc}A.
\]
Therefore
\[
A\gamma^\mu A
=
\frac{2p^\mu}{mc}A-\gamma^\mu.
\]
Substituting this back,
\[
\Pi_s\gamma^\mu\Pi_s
=
\frac{1}{4}
\left(
\gamma^\mu
+
s\frac{2p^\mu}{mc}1_4
+
\frac{2p^\mu}{mc}A
-
\gamma^\mu
\right).
\]
The \(\gamma^\mu\) terms cancel, leaving
\[
\Pi_s\gamma^\mu\Pi_s
=
\frac{p^\mu}{2mc}
\left(
s1_4+A
\right).
\]
Since
\[
s\Pi_s
=
\frac{1}{2}
\left(
s1_4+A
\right),
\]
we obtain
\[
\Pi_s\gamma^\mu\Pi_s
=
s\frac{p^\mu}{mc}\Pi_s.
\]

\bibliography{ref}
\end{document}